\documentclass[aps,prd,preprint,eqsecnum,floats,epsf,superscriptaddress,nofootinbib]{revtex4}
\usepackage{epsfig}
\usepackage{graphicx}% Include figure files
\usepackage{extarrows}
\usepackage{subfigure}
\usepackage{color}

\def\ol{\overline}
\def\ov{\overline}
\def\be{\begin{eqnarray}}
\def\en{\end{eqnarray}}
\def\non{\nonumber}
\def\CP{{\it CP}~}

\def\la{\langle}
\def\ra{\rangle}

\def\B{{\cal B}}
\def\M{{\cal M}}

\def\PAP{{P\!A_P}}
\def\PAV{{P\!A_V}}
\def\PEP{{P\!E_P}}
\def\PEV{{P\!E_V}}
\def\PE{{P\!E}}
\def\PA{{P\!A}}
\def\lsim{ {\ \lower-1.2pt\vbox{\hbox{\rlap{$<$}\lower5pt\vbox{\hbox{$\sim$}
}}}\ } }
\def\gsim{ {\ \lower-1.2pt\vbox{\hbox{\rlap{$>$}\lower5pt\vbox{\hbox{$\sim$}
}}}\ } }

\begin{document}

\font\el=cmbx10 scaled \magstep2{\obeylines\hfill July, 2021}
\vskip 1.0 cm

\title{{\it CP} violation in quasi-two-body $D\to V\!P$ decays and three-body $D$ decays mediated by vector resonances}

\author{Hai-Yang Cheng}
\affiliation{Institute of Physics, Academia Sinica, Taipei, Taiwan 11529, ROC}

\author{Cheng-Wei Chiang}
\affiliation{Department of Physics and Center for Theoretical Physics, National Taiwan University, Taipei, Taiwan 10617, ROC}
\affiliation{Physics Division, National Center for Theoretical Sciences, Taipei, Taiwan 10617, ROC}

%\date{\today}

\begin{abstract}
\bigskip

We re-examine direct \CP violation in the quasi-two-body $D\to V\!P$ decays and study \CP asymmetries in three-body $D$ decays proceeding through intermediate vector resonances within the framework of the topological amplitude approach for tree amplitudes and the QCD factorization approach for penguin amplitudes.  Using the same mechanism of incorporating long-distance penguin-exchange amplitude induced from final-state rescattering that nicely accounts for the \CP asymmetry difference between $D^0\to K^+K^-$ and $\pi^+\pi^-$ modes, we find that \CP asymmetry can also occur at the per mille level in many of the $D\to V\!P$ channels or otherwise be negligibly small.  We point out six golden modes which have sufficiently large branching fractions and direct \CP violation of order $10^{-3}$.  In particular, the direct \CP asymmetry difference $\Delta a_{CP}^{V\!P}$ between $D^0\to K^+K^{*-}$ and $\pi^+\rho^-$ is predicted to be $(-1.61\pm0.33)\times 10^{-3}$, very similar to the counterpart in the $P\!P$ sector.  Besides, we find a positive and large \CP asymmetry of order $10^{-3}$ in $D^0\to K_S K^{*0}$, an example of the asymmetry induced at the tree level.  We take into account the flavor-singlet QCD-penguin contributions to calculate their effects on the modes such as $D\to\pi\phi$ and $D\to \eta \phi$.  We compare our results with the other approach in detail.  For three-body $D$ decays, we perform the Dalitz plot analysis of those decays mediated by several vector resonances with the same mass but different electric charges.  Local \CP asymmetry varies in magnitude and sign from region to region and can reach the percent level in certain local regions of phase space due to interference.

\end{abstract}

%\end{abstract}


\maketitle
\small

%\tableofcontents
\newpage

%%%%%%%%%%%%%%%%%%%%%%%%%%%%%%%%%%%%%%%%%%%%%%%%%%
\section{Introduction \label{sec:intro}}
%%%%%%%%%%%%%%%%%%%%%%%%%%%%%%%%%%%%%%%%%%%%%%%%%%
In 2019  the LHCb Collaboration has announced the measured difference between the time-integrated \CP asymmetries of the decays $D^0\to K^+K^-$ and $D^0\to\pi^+\pi^-$ based on pion and muon tagged analyses~\cite{LHCb:2019}. Combining these with previous LHCb results in 2014 \cite{LHCb:2014} and 2016 \cite{LHCb:2016} leads to a result of a nonzero value of $\Delta A_{CP}$~\cite{LHCb:2019}
\be \label{eq:LHCb:2019}
\Delta A_{CP}\equiv A_{CP}(K^+K^-)-A_{CP}(\pi^+\pi^-)=(-1.54\pm0.29)\times 10^{-3}.
\en
The time-integrated asymmetry can be further decomposed into a direct \CP asymmetry $a_{CP}^{\rm dir}$
and a mixing-induced indirect \CP asymmetry characterized by the parameter $A_\Gamma$ which measures the asymmetry between $D^0\to f$ and $\bar D^0\to f$ effective decay widths
\be
A_{CP}(f)\approx a_{CP}^{\rm dir}(f) -{\la t\ra\over\tau(D^0)} A_\Gamma(f) \ ,
\en
where  $\la t\ra$ is the average decay time in the sample. To a good approximation, $A_\Gamma$ is independent of the decay mode and hence,
\be
\Delta A_{CP}=\Delta a_{CP}^{\rm dir} -{\Delta \la t\ra\over\tau(D^0)} A_\Gamma \ ,
\en
where $\Delta \la t\ra$ is the difference in the mean decay times of the two decays.
Based on the LHCb average of $A_\Gamma$, it follows that the direct \CP asymmetry difference is given by \cite{LHCb:2019}
\be
\Delta a_{CP}^{\rm dir}=(-1.57\pm0.29)\times 10^{-3}.
\en
This is the first observation of \CP violation at $5.3\sigma$ in the charm sector.

The most important question to be addressed is whether or not the observed \CP violation in the charm sector  is consistent with the standard model.
This is the main issue discussed in most of recent papers appearing after the 2019 LHCb measurement~\cite{Xing:2019uzz,Chala:2019fdb,Li:2019hho,Grossman:2019xcj,Soni:2019xko,Cheng:2019,Nir,Calibbi:2019bay,Soni:2020,Bigi,Saur:2020rgd,Bediaga:2020qxg,Lenz:2020awd,Dery:2021mll}.
\footnote{Various models including many new physics ones (see, for example, references [2-23] listed in~\cite{Cheng:2019}) had been proposed in the past to explain the large $\Delta A_{C\!P}=-(0.82\pm0.21\pm0.11)\%$ reported by the LHCb in 2012~\cite{LHCb2012}. However, this large $\Delta A_{C\!P}$ was not seen anymore in subsequent LHCb analyses~\cite{LHCb:2013,LHCb:2014,LHCb:2016}.  }
One may na{\"i}vely argue that it is difficult to accommodate the LHCb's measurement of \CP asymmetry within the standard model. To see this, consider the tree $T$ and penguin $P$ contributions to $D^0\to K^+K^-$ and $D^0\to \pi^+\pi^-$. Their interference gives rise to $\Delta a_{CP}^{\rm dir}$. A simplified expression of the \CP asymmetry difference between them is given by
\footnote{  A complete expression of $\Delta a_{CP}^{\rm dir}$ will be given in Eq.~\eqref{eq:DCPVdiff} below. }
\be
\Delta a_{CP}^{\rm dir}\approx -1.3\times 10^{-3} \left( \left|{P\over T}\right|_{_{K\!K}}\sin\theta_{_{K\!K}}+ \left|{P\over T}\right|_{_{\pi\pi}}\sin\theta_{_{\pi\pi}} \right),
\en
where the factor of $1.3\times 10^{-3}$ comes from the imaginary part of a certain combination of CKM matrix elements, $2{\rm Im}(\lambda_d\lambda_s^*)/|\lambda_d|^2$, with $\lambda_p\equiv V^*_{cp}V_{up}$ for $p=d,s,b$, and
$\theta{_{K\!K}}$ is the strong phase of $(P/T)_{_{K\!K}}$ and likewise for $\theta{_{\pi\pi}}$.  Since $|P/T|$ is na{\"i}vely expected to be of order $(\alpha_s(\mu_c)/\pi)\sim {\cal O}(0.1)$, it appears that $\Delta a_{CP}^{\rm dir}$ is most likely of order $10^{-4}$ even if the strong phases are allowed to be close to $90^\circ$. Indeed, using the results of $|P/T|$ obtained from light-cone sum rules ~\cite{Khodjamirian:2017}, 
\be
\left|{P\over T}\right|_{\pi\pi}=0.093\pm0.011, \qquad \left|{P\over T}\right|_{_{KK}}=0.075\pm0.015,
\en
Khodjamirian and Petrov  argued an upper bound in the standard model, namely, $|\Delta a_{CP}^{\rm SM}|\leq (2.0\pm0.3)\times 10^{-4}$.

Before the LHCb observation of $\Delta A_{C\!P}$ in 2019, there existed two independent studies in 2012 in which direct \CP violation in charmed meson decays was explored based on the topological diagram approach for tree amplitudes and the QCD-inspired approach for penguin amplitudes~\cite{Cheng:2012b,Cheng:2012a,Li:2012}. Interestingly, both works predicted a $\Delta A_{C\!P}$ of order 0.1\% with a negative sign seven years before the LHCb observation of
\CP asymmetry in the charm sector.
It followed that
\be \label{eq:PoverT QCDF}
\left({P\over T}\right)_{\pi\pi}\approx 0.23 e^{-i150^\circ}, \qquad \left({P\over T}\right)_{K\!K}\approx 0.22 e^{-i150^\circ},
\en
were found in the QCD factorization (QCDF) calculation~\cite{Cheng:2019,Cheng:2012b,Cheng:2012a}, while
\be
\left({P\over T}\right)_{\pi\pi}\approx 0.30 e^{i110^\circ}, \qquad \left({P\over T}\right)_{K\!K}\approx 0.24 e^{i110^\circ},
\en
were obtained in the so-called factorization-assisted topological amplitude (FAT) approach
~\cite{Li:2012}.
Contrary to the sum-rule calculation, the magnitude of $P/T$ turns out to be of order $(0.22\sim 0.30)$ in the QCD-inspired approaches. However, it does not suffice to explain the measured value of $\Delta a_{\rm CP}^{\rm dir}$.

The complete expression of $\Delta a_{\rm CP}^{\rm dir}$ is given by~\cite{Cheng:2012b}
\be \label{eq:DCPVdiff}
\Delta a_{\CP}^{\rm dir}=-1.30\times 10^{-3} \left( \left| {P^d+P\!E^d+P\!A^d \over T+E^s-\Delta P}\right|_{_{K\!K}}\sin\delta_{_{K\!K}}+\left| {P^s+P\!E^s+P\!A^s\over T+E^d+\Delta P}\right|_{\pi\pi}\sin\delta_{\pi\pi}\right),
\en
where $E$ is the $W$-exchange amplitude, $\PE$ is the QCD-penguin exchange amplitude and $\PA$ the QCD-penguin annihilation amplitude, the superscript $d$ or $s$ refers to the quark involved in the associated penguin loop, and $\delta_{_{K\!K}}$ is the strong phase of $(P^d+P\!E^d+P\!A^d)_{_{K\!K}}$ relative to $(T+E^s-\Delta P)_{_{K\!K}}$ and likewise for the phase $\delta_{\pi\pi}$. The parameter $\Delta P$ is defined by $\Delta P \equiv (P^d+\PE^d+\PA^d)-(P^s+\PE^s+\PA^s)$. The observation of the decay $D^0\to K^0\ov K^0$ indicates that SU(3) symmetry must be broken in the topological amplitude $E$. A fit to the data yields two possible solutions~\cite{Cheng:2019}
\be \label{eq:EdEs}
\begin{split}
{\rm I: }~~ &  E^d=1.10\, e^{i15.1^\circ}E ~, \qquad E^s=0.62\, e^{-i19.7^\circ}E
~; \\
{\rm II: }~~ &  E^d=1.10\, e^{i15.1^\circ}E ~, \qquad E^s=1.42\, e^{-i13.5^\circ}E
~,
\end{split}
\en
where $E^q$ refers to the $W$-exchange amplitude associated with $c\bar u\to q\bar q$ ($q=d,s$).

It follows that
\be \label{eq:Ratio,QCDF}
\left({P^s+P\!E^s+P\!A^s \over T+E^d +\Delta P}\right)_{\pi\pi}=0.40\, e^{i176^\circ}, \quad
\left({P^d+P\!E^d+P\!A^d \over T+E^s-\Delta P }\right)_{_{K\!K}}=
\begin{cases}
0.29\,e^{-i164^\circ} \\
0.29\,e^{i178^\circ}
\end{cases},
\en
in QCDF for Solutions I and II of $W$-exchange amplitudes $E^d$ and $E^s$~\cite{Cheng:2019}, and
\be \label{eq:Ratio,pQCD}
\left({P^s+P\!E^s+P\!A^s \over T+E^d +\Delta P}\right)_{\pi\pi}=0.66\, e^{i134^\circ}, \qquad
\left({P^d+P\!E^d+P\!A^d \over T+E^s-\Delta P }\right)_{_{K\!K}}= 0.45\, e^{i131^\circ}
\en
in FAT~\cite{Li:2012}. Now we see a crucial difference between the two calculations.
The calculated phases $\delta_{\pi\pi}$ and $\delta_{K\!K}$ are close to $180^\circ$ in QCDF, while they deviate from $90^\circ$ by only around $40^\circ$ in FAT. As a consequence,  Eq.~(\ref{eq:Ratio,pQCD}) leads to $\Delta a_{\rm CP}^{\rm dir}\approx -1.0\times 10^{-3}$, but not so for QCDF, where $\Delta a_{\rm CP}^{\rm dir}\approx 6.8\times 10^{-5}$ and $-4.9\times 10^{-5}$ for Solutions I and II of $W$-exchange amplitudes, respectively.

One of the most salient features of the topological approach for charm decays
is that all the topological amplitudes except the tree amplitude $T$
are dominated by nonfactorizable long-distance effects. That is, final-state interactions play an essential role in describing the hadronic decays of charmed mesons. In~\cite{Cheng:2012a}
we have pointed out the importance of a resonant-like final-state rescattering which has the same topology as the QCD-penguin exchange topological graph $\PE$. Hence, penguin annihilation receives sizable long-distance contributions through final-state interactions. Making the ansatz that $(\PE)^{\rm LD}$ is of the same order of magnitude as $E$ and flavor independent, we obtain
\be \label{eq:PoverT+E}
\begin{split}
\left({P^s+P\!E^s+P\!A^s+\PE^{\rm LD} \over T+E^d +\Delta P}\right)_{\pi\pi} &= 0.81\, e^{i119^\circ}
~, \\
\left({P^d+P\!E^d+P\!A^d+\PE^{\rm LD} \over T+E^s-\Delta P }\right)_{_{K\!K}} &=
\begin{cases}
0.48\,e^{i143^\circ} \\
0.48\,e^{i126^\circ}
\end{cases}.
\end{split}
\en
Comparing this with Eq.~(\ref{eq:Ratio,QCDF}), we see that, in the presence of the long-distance penguin exchange,  not only the magnitudes of the ratios are enhanced but the strong phases are also closer to $90^\circ$ rather than to $180^\circ$.
We have predicted that $\Delta a_{CP}^{\rm dir}$ is about $(-0.139\pm0.004)\%$ and $(-0.151\pm0.004)\%$ for the two solutions of $W$-exchange amplitudes~\cite{Cheng:2012b}. Those were the main predictions among others made in 2012.  Nowadays, we know that the LHCb's new measurement (\ref{eq:LHCb:2019}) almost coincides with our second solution.
This implies that one does not need new physics at all to understand the first observation of $\Delta a_{CP}^{\rm dir}$ by the LHCb.

Although both QCDF and FAT approaches lead to $\Delta a_{CP}^{\rm dir}$ at the per mille level~\cite{Cheng:2012b,Cheng:2012a,Li:2012}, they can be discriminated in the $D\to V\!P$ sector. While our approach based on QCDF predicts \CP asymmetries of $D^0\to \pi^+\rho^-$ and $D^0\to K^+K^{*-}$ similar to that of the corresponding $D\to PP$ decays~\cite{Cheng:2019}, the FAT approach leads to very tiny values of \CP violation for $D^0\to \pi^\pm\rho^\mp$ and $D^0\to K^\pm K^{*\mp}$~\cite{Qin}. In this work, we shall examine the underlying reason for the discrepancy between the two approaches for $D\to V\!P$ decays and present a more detailed analysis. We will also study the implications for \CP asymmetry distributions in three-body $D$ decays. For example, the experimental Dalitz plot analysis of $D^0\to \pi^+\pi^-\pi^0$ indicates that this channel is predominated by the $\rho^\pm$ and $\rho^0$ resonances  (see Sec.~III B below). The interference between these resonances will yield rich information on local \CP asymmetries.

The layout of the present work is organized as follows.  In view of updates in several branching fractions, we will perform a re-fit to the Cabibbo-favored modes to extract various topological tree amplitudes in Sec.~\ref{sec:DCPV}.  Then we elaborate the calculation of direct \CP violation in $D\to V\!P$ decays in detail and compare our results with those of FAT. In Sec.~\ref{sec:3-body DCPV} we discuss \CP asymmetry distributions in the following 3-body $D$ decays: $D^0\to K^+K^-\pi^0$, $D^0\to \pi^+\pi^-\pi^0$, $D^+\to K^+K_S\pi^0$ and $D_s^+\to K^0\pi^+\pi^0$. We focus on the phase-space region where these three-body decays
are governed by the quasi-two-body $D\to V\!P$ ones.
Our conclusions are presented in Sec.~\ref{sec:summary}.

%%%%%%%%%%%%%%%%%%%%%%%%%%%%%%%%%%%%%%%%%%%%%%%%%%
\section{Direct \CP violation in $D\to V\!P$ decays \label{sec:DCPV}}
%%%%%%%%%%%%%%%%%%%%%%%%%%%%%%%%%%%%%%%%%%%%%%%%%%

\subsection{Topological tree amplitudes \label{sec:Topological tree amplitudes}}
For $D\to V\!P$ decays, there exist two different types of topological diagrams since the spectator quark of the charmed meson may end up in the pseudoscalar or vector meson. For topological color-allowed tree amplitude $T$ and color-suppressed amplitude $C$ in $D\to V\!P$ decays, the subscript $P$ ($V$) implies that the pseudoscalar (vector) meson contains the spectator quark of the charmed meson.  For the $W$-exchange amplitude $E$ and $W$-annihilation $A$ with the final state $q_1\bar q_2$, the subscript $P$ ($V$) denotes that the pseudoscalar (vector) meson contains the antiquark $\bar q_2$.

The partial decay width of the $D$ meson into a vector and a pseudoscalar mesons is given by
\begin{equation}
\Gamma(D\to V\!P)=\frac{p_c}{8\pi m_D^2}\sum_{\rm pol}|{{\cal  M}}|^2,
\label{decaywidthC}
\end{equation}
with $p_c$ being the center-of-mass (c.m.) momentum of either final-state particle.
This expression can be simplified by the replacement $\epsilon\cdot p_D\to p_c(m_D/m_V)$ after the polarization states of the vector meson are summed over. Hence,
\begin{equation}
\Gamma(D\to V\!P)=\frac{p_c^3}{8\pi m_V^2}|\tilde{{\cal M}}|^2 ~,
\label{decaywidthA}
\end{equation}
with ${\cal M}=\tilde{\cal M}(\epsilon\cdot p_D)$.
By performing a $\chi^2$ fit to the Cabibbo-favored (CF) $D\to V\!P$ decays, we have extracted the magnitudes and strong phases of the topological amplitudes $T_V,C_V,E_V,A_V$ and $T_P,C_P,E_P,A_P$ from the measured partial widths through Eq.~(\ref{decaywidthA}) and found many possible solutions with local $\chi^2$ minima~\cite{Cheng:2016,Cheng:2019}.  Previously, we found six best $\chi^2$-fit solutions (S1)--(S6) as listed in Table~IV of~\cite{Cheng:2019}, when we restricted ourselves to $\chi^2_{\rm min} < 10$.
Although Solutions (S1)--(S6) generally fit the CF modes well, they led to very different predictions for some of the singly Cabibbo-suppressed (SCS) decays.
Especially, the $D^0\to\pi^0\omega$, $D^+\to \pi^+\rho^0$ and $D^+\to\pi^+\omega$ decays were very useful in discriminating among different solutions. For example, Solutions (S1), (S2), (S4) and (S5) implied that the branching fraction of $D^+\to \pi^+\rho^0$ should be smaller than that of  $D^+\to \pi^+\omega$, in contradiction with the experimental data.
As shown in~\cite{Cheng:2019}, Solution (S6) gave a better description of the SCS modes.

In light of the recent new measurements of $D^0\to \ov K^{*0}\eta$ by Belle~\cite{Bell:Dto Ksteta}, $D^0\to \ov K^0\phi$, $D_s^+\to \ov K^{*0}K^+,\pi^+\phi$ by BESIII~\cite{BESIII:DtoKphi,BESIII:Ds KstK}, we have performed a new fit to the data given in Table~\ref{table:VPCF} and now find five solutions (S1')--(S5'), listed in Table~\ref{tab:CFVPB}.  It is noted that there is a sign ambiguity if all the strong phases change their sign simultaneously.  The topological amplitudes of all these solutions respect the hierarchy pattern:
\be \label{eq:tree_hierarchy}
|T_P|>|T_V|\gsim|C_{P}|>|C_{V}|\gsim|E_P|>|E_V|>|A_{P,V}|.
\en
Among them, Solution (S3') is very similar to Solution (S6) and provides a best description of the SCS modes. The topological amplitude sizes and the strong phases for this solution are presented
in Table~\ref{tab:CFVPB}, where we have chosen the $\eta-\eta'$ mixing angle $\phi=43.5^\circ$~\cite{Aaij:2014jna}, which is defined in the flavor basis
\be
 \begin{pmatrix} \eta \cr \eta' \end{pmatrix}
 = \begin{pmatrix} \cos\phi & -\sin\phi \cr \sin\phi & \cos\phi\cr \end{pmatrix}
 \begin{pmatrix} \eta_q \cr \eta_s \end{pmatrix},
\en
with $\eta_q={1\over\sqrt{2}}(u\bar u+d\bar d)$ and $\eta_s=s\bar s$.  Note that (S3') in Table~\ref{tab:CFVPB} here and (S6) in Table IV of~\cite{Cheng:2019} are quite similar except for $E_V$ and the phase of $T_P$. It is not surprising that the magnitude of $E_V(S3')=0.58\pm0.06$ is about twice bigger than that of $E_V(S6)=0.26\pm0.04$~\cite{Cheng:2019} because the new average of $\B(D^0\to \ov K^{*0}\eta)=(1.352\pm0.115)\%$ is increased by 35\% from the old one $(1.02\pm0.30)\%$~\cite{PDG}, recalling that its amplitude is proportional to $\left[ {1\over\sqrt{2}}(C_P+E_P)\cos\phi-E_V\sin\phi \right]$. As a consequence, the phase of $E_V$ is also changed significantly from $(224^{+22}_{-40})^\circ$~\cite{Cheng:2019} to $(283\pm5)^\circ$ with a much improved accuracy.
Since $A(D^0\to K^-\rho^+)\propto (T_P+E_V)$, $T_P$ is affected by a change of $E_V$ accordingly.

%%%%%%%%%%%%%%%%%%%%
%\squeezetable
%\renewcommand{\arraystretch}{0.85}
\begin{table}[tp!]
\caption{Flavor amplitude decompositions, experimental branching fractions, and predicted branching fractions for the Cabibbo-favored $D \to VP$ decays.  Data are taken from the Particle Data Group~\cite{PDG}, Belle~\cite{Bell:Dto Ksteta} and BESIII~\cite{BESIII:DtoKphi,BESIII:Ds KstK}.
Here  $\lambda_{sd}\equiv V_{cs}^*V_{ud}$.  The column of ${\cal B}_{\rm theory}(S3')$ shows predictions based on Solutions (S3') presented in Table~\ref{tab:CFVPB}.  All branching fractions are quoted in units of \%.  }
\vspace{6pt}
\begin{ruledtabular}
\begin{tabular}{l l l c c c c c }
Meson & Mode & Amplitude decomposition
& ${\cal B}_{\rm exp}$ & ${\cal B}_{\rm theory}(S3')$
\\
\hline
$D^0$  & $K^{*-}\,\pi^+$          & $\lambda_{sd}(T_V + E_P)$
&$5.34\pm0.41$                    &$5.41\pm0.33$		
\\
&$K^-\,\rho^+$                      & $\lambda_{sd}(T_P + E_V)$
&$11.3 \pm 0.7$                    &$11.4\pm0.6$		
\\
& $\ol{K}^{*0}\,\pi^0$             &$\frac{1}{\sqrt{2}}\lambda_{sd}(C_P - E_P)$
&$3.74\pm0.27$                     &$3.64\pm0.17$	
\\
& $\ol{K}^0\,\rho^0$              & $\frac{1}{\sqrt{2}}\lambda_{sd}(C_V - E_V)$
&$1.26^{+0.12}_{-0.16}$           &$1.30\pm0.12$	
\\
& $\ol{K}^{*0}\,\eta$               & $\lambda_{sd}\left[ {1\over \sqrt{2}}(C_P + E_P)\cos\phi - E_V\sin\phi\, \right]$
&$1.352\pm0.115$\footnotemark[1]                     &$1.33\pm0.08$		
\\
& $\ol{K}^{*0}\,\eta\,'$            &$-\lambda_{sd}\left[ {1\over \sqrt{2}}(C_P + E_P)\sin\phi + E_V\cos\phi\, \right]$
&$<0.10$                            &$0.0034\pm0.0003$
\\
& $\ol{K}^0\,\omega$             & $-\frac{1}{\sqrt{2}}\lambda_{sd}(C_V + E_V)$
&$2.22 \pm 0.12$                 &$2.24\pm0.16$	
\\
& $\ol{K}^0\,\phi$                   &$-\lambda_{sd}E_P$
&$0.866\pm0.029$\footnotemark[2]   			& $0.868\pm0.028$		
\\
\hline
$D^+$
& $\ol{K}^{*0}\,\pi^+$  & $\lambda_{sd}(T_V + C_P)$
&$1.57\pm0.13$                   &$1.58\pm0.13$		
\\
&$\ol{K}^0\,\rho^+$               & $\lambda_{sd}(T_P + C_V)$
&$12.3^{+1.2}_{-0.7}$             &$12.5\pm0.4$		
\\
\hline
$D_s^+$
& $\ol{K}^{*0}\,K^+$& $\lambda_{sd}(C_P + A_V)$
&$3.91\pm0.09$\footnotemark[3]                    &$3.92\pm0.13$	
\\
&$\ol{K}^0\,K^{*+}$              & $\lambda_{sd}(C_V + A_P)$
&$5.4 \pm 1.2$                    &$2.95\pm0.20$		
\\
&$\rho^+\,\pi^0$                   & $\frac{1}{\sqrt{2}}\lambda_{sd}(A_P - A_V)$
&---                                  &$0.023\pm0.014$		
\\
&$\rho^+\,\eta$                    & $\lambda_{sd}\left[ {1\over\sqrt{2}}( A_P + A_V)\cos\phi-T_P \sin\phi \right]$
&$8.9 \pm 0.8$                     &$8.95\pm0.36$
\\
&$\rho^+\,\eta\,'$                 & $\lambda_{sd} \left[{1\over\sqrt{2}}( A_P + A_V)\sin\phi+T_P\cos\phi \right]$
&$5.8\pm1.5$   						&$3.43\pm0.12$	
\\
& $\pi^+\,\rho^0$                 & $\frac{1}{\sqrt{2}}\lambda_{sd}(A_V - A_P)$
 &$0.019\pm0.012$               &$0.023\pm0.014$	
\\
& $\pi^+\,\omega$                &$\frac{1}{\sqrt{2}}\lambda_{sd}(A_V + A_P)$
&$0.192\pm 0.030$                 &$0.19\pm0.04$	
\\
& $\pi^+\,\phi$                      &$\lambda_{sd}T_V$
&$4.53 \pm 0.12$\footnotemark[3]                      &$4.52\pm0.11$	
\\
\end{tabular}
\label{table:VPCF}
\footnotetext[1]{The new measurement from Belle $\B(D^0\to \ov K^{*0}\eta)=(1.41^{+0.13}_{-0.12})\%$ ~\cite{Bell:Dto Ksteta} is taken into account in the world average.}
\footnotetext[2]{The new measurement from BESIII~\cite{BESIII:DtoKphi}  is taken into account in the world average.}
\footnotetext[3]{The new measurement from BESIII~\cite{BESIII:Ds KstK}  is taken into account in the world average.}
\end{ruledtabular}
\end{table}
%%%%%%%%%%%%%%%%%%%%

%%%%%%%%%%%%%%%%%%%%
\begin{table}[t]
\caption{Solutions with $\chi^2_{\rm min} \le 10$ obtained using Eq.~(\ref{decaywidthA}) and $\phi=43.5^\circ$.  We will take Solution (S3') in subsequent analyses.  The amplitude sizes are quoted in units of  $10^{-6}(\epsilon\cdot p_D)$ and the strong phases in units of degrees.
}
\label{tab:CFVPB}
\medskip
\footnotesize{
\begin{ruledtabular}
\begin{tabular}{c c c c c c c c c c c}
Set       &$|T_V|$                   &$|T_P|$            &$\delta_{T_P}$          &$|C_V|$                          &$\delta_{C_V}$          &$|C_P|$                  &$\delta_{C_P}$                &$|E_V|$                         &$\delta_{E_V}$     \\
          &$|E_P|$                            &$\delta_{E_P}$   &$|A_V|$                           &$\delta_{A_V}$             &$|A_P|$ &$\delta_{A_P}$ &$\chi^2_{\rm min}$ &fit quality\\
\hline
(S1')  &$2.19\pm0.03$ &$3.32\pm0.06$ &$100\pm3$ &$1.75\pm0.04$ &$312\pm3$ &$2.10\pm0.02$ &$201\pm1$ &$0.29\pm0.04$ &$334^{+14}_{-22}$  \\
            &$1.69\pm0.03$ &$109\pm2$ &$0.20\pm0.02$ &$32^{+8}_{-11}$ &$0.21\pm0.03$  &$357^{+12}_{-9}$           & 5.88 & 11.74\% \\ %GS'4
\hline
(S2')  &$2.19\pm0.03$ &$3.23\pm0.06$ &$264\pm3$ &$1.74\pm0.04$ &$50\pm3$ &$2.07\pm0.02$ &$201\pm1$ &$0.36\pm0.05$ &$8^{+9}_{-8}$  \\
            &$1.69\pm0.03$ &$109\pm2$ &$0.20\pm0.02$ &$349^{+9}_{-7}$ &$0.22\pm0.03$  &$23^{+9}_{-12}$           & 6.39 & 9.41\% \\ %GS'4
\hline
(S3')  &$2.19\pm0.03$ &$3.56\pm0.06$ &$61\pm5$ &$1.69\pm0.04$ &$220\pm3$ &$2.02\pm0.02$ &$201\pm1$    &$0.58\pm0.06$ &$283\pm5$  \\
            &$1.69\pm0.03$ &$108\pm3$ &$0.23\pm0.02$ &$77\pm5$ &$0.18\pm0.03$  &$111^{+13}_{-10}$           & 7.06 & 6.99\% \\ %GS'4
\hline
(S4')  &$2.19\pm0.03$ &$3.50\pm0.06$ &$106^{+3}_{-4}$ &$1.74\pm0.04$ &$262^{+4}_{-3}$ &$2.09\pm0.02$ &$201\pm1$ &$0.38\pm0.05$ &$308^{+8}_{-10}$  \\
            &$1.69\pm0.03$ &$109\pm2$ &$0.25\pm0.02$ &$62^{+6}_{-7}$ &$0.14\pm0.03$  &$44^{+50}_{-16}$           & 7.34 & 6.19\% \\ %GS'4
\hline
(S5')  &$2.19\pm0.03$ &$3.54\pm0.06$ &$268^{+5}_{-4}$ &$1.67\pm0.04$ &$107^{+3}_{-4}$ &$2.04\pm0.02$ &$201\pm1$ &$0.62^{+0.06}_{-0.07}$ &$43\pm4$  \\
            &$1.69\pm0.03$ &$108\pm2$ &$0.26\pm0.02$ &$324\pm5$ &$0.13\pm0.02$  &$329^{+24}_{-32}$           & 8.57 & 3.56\% \\ %GS'4
\end{tabular}
\end{ruledtabular}
\label{fitting}}
\end{table}
%%%%%%%%%%%%%%%%%%%%

%%%%%%%%%%%%%%%%%%%%%%%%%%%%%%%%%%%
\begin{table}[h]
\caption{Solutions of the parameters $e_{V,P}^{d,s}$ and the phases $\delta e_{V,P}^{d,s}$ (in units of degrees) describing SU(3) breaking effects in the $W$-exchange amplitudes.
  \label{tab:EPEV}}
  \medskip
\begin{ruledtabular}
\begin{tabular}{ l c  c c c c c c c}
 &  $e_V^d$ & $\delta e_V^d$ & $e_P^d$ & $\delta e_P^d$ & $e_V^s$ & $\delta e_V^s$ & $e_P^s$ & $\delta e_P^s$  \\
\hline
(i) & 0.34 & 347 & 0.62 & 343 & 0.24 & 101 & 0.62 & 149 \\
(ii) & 0.34 & 347 & 0.62 & 343 & 0.24 & 101 & 0.40 & 349 \\
(iii) & 0.34 & 347 & 0.62 & 343 & 1.63 & 229 & 0.62 & 149 \\
(iv) & 0.34 & 347 & 0.62 & 343 & 1.63 & 229 & 0.40 & 349 \\
\end{tabular}
\end{ruledtabular}
\end{table}
%%%%%%%%%%%%%%%%%%%%%%%%%%%%%%%%%

In the $P\!P$ sector, we need SU(3) breaking in the $W$-exchange diagrams [see Eq.~(\ref{eq:EdEs})] in order to induce the observed $D^0\to K_SK_S$ decay and to explain the large rate difference between $D^0\to K^+K^-$ and $D^0\to \pi^+\pi^-$ decays.
Likewise, we also need SU(3) breaking in the $W$-exchange amplitudes in the $V\!P$ sector because (a) the ratios, $\Gamma(D^0\to K^+K^{*-})/\Gamma(D^0\to \pi^+\rho^-)=0.32\pm0.03$ and $\Gamma(D^0\to K^-K^{*+})/\Gamma(D^0\to \pi^-\rho^+)=0.45\pm0.03$~\cite{PDG}, deviate substantially from unity expected in the SU(3) limit and (b) the predicted rates of $D^0\to K^0\ol K^{*0}$ and $D^0\to \ol K^0 K^{*0}$ modes are too large~\cite{Cheng:2019}. Writing
\be \label{eq:SU(3)inW}
\begin{split}
E_V^d=e_V^d e^{i\delta e_V^d}E_V, \quad E_V^s=e_V^s e^{i\delta e_V^s}E_V,
\\
E_P^d=e_P^d e^{i\delta e_P^d}E_P, \quad E_P^s=e_P^se^{i\delta e_P^s} E_P,
\end{split}
\en
we are able to determine the eight unknown parameters $e_V^d,e_P^d,e_V^s,e_P^s$ and $\delta e_V^d,\delta e_P^d,\delta e_V^s,\delta e_P^s$ from the branching fractions of the following eight modes: $D^0\to \pi^+\rho^-,\pi^-\rho^+,\pi^0\rho^0,\pi^0\omega$ and $D^0\to K^+K^{*-}.K^-K^{*+},K^0\ol K^{*0},\ol K^0K^{*0}$.
Table~\ref{tab:EPEV} shows the four solutions of SU(3) breaking effects in the $W$-exchange amplitudes for Solution (S3').  Recall that in our previous work, there are six solutions for the parameters $e_{V,P}^{d,s}$ and the associated phases (see Table~VII of~\cite{Cheng:2019}).
As we shall show in Sec.~\ref{sec:Branching fractions and SU(3) breaking} below, the four different solutions of SU(3) breaking in $W$-exchange amplitudes can be discriminated using the SCS mode $D^0\to \eta\phi$.  It turns out that Solution (iv) is preferred.

%%%%%%%%%%%%%%%%%%%%%%%%%%%%%%%%%%%%%%%%%%%%%%%%%%
\subsection{Penguin amplitudes in QCD factorization \label{sec:penguin}}
%%%%%%%%%%%%%%%%%%%%%%%%%%%%%%%%%%%%%%%%%%%%%%%%%%

Although the topological tree amplitudes $T,C,E$ and $A$ for two-body hadronic $D$ decays can be extracted from the data, information on penguin amplitudes (QCD penguin, penguin annihilation, etc.) needed for inducing \CP violation in the SCS decays is still absent. To consider the penguin contributions,
we shall take the QCD factorization (QCDF) approach~\cite{BBNS99,BN} to evaluate the hadronic matrix elements.
Although the QCD-inspired approaches such as QCDF and pQCD~\cite{Keum:2000wi} have been applied rather successfully to describing weak hadronic decays of $B$ mesons, it should be kept in mind that these approaches provide only a crude estimate of the penguin amplitudes in charm decays because the charm quark mass is not heavy enough and $1/m_c$ power corrections are so large that a sensible heavy quark expansion is not allowed.

%%%%%%%%%%%%%%%%%%%%%%%%%%%%%%%%%%%
\begin{figure}[t]
\vspace*{1ex}
\includegraphics[width=1.5in]{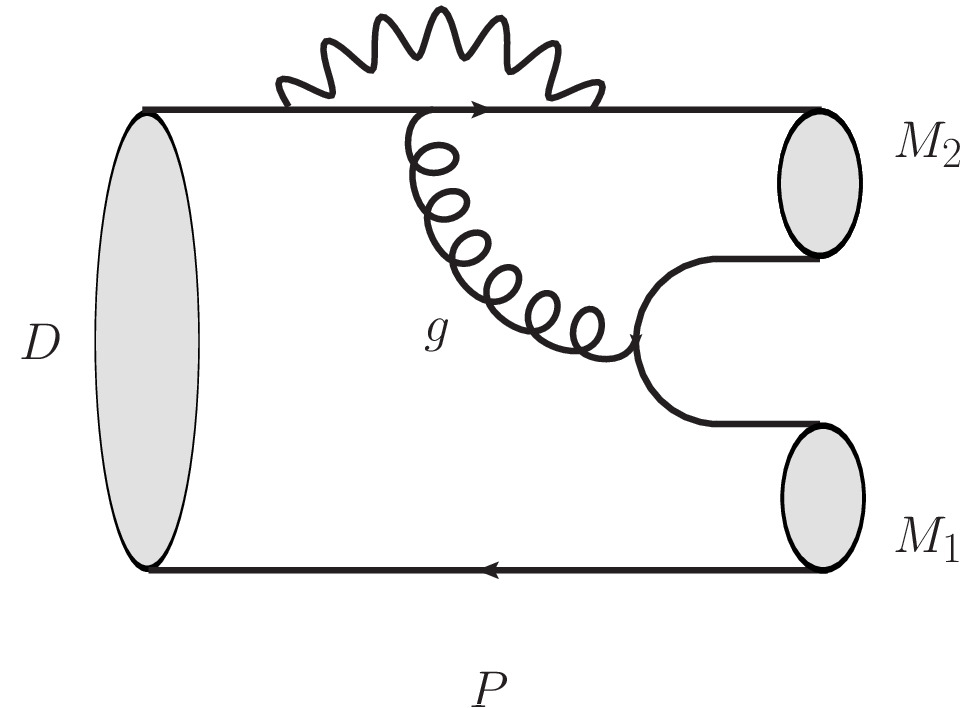} \includegraphics[width=1.5in]{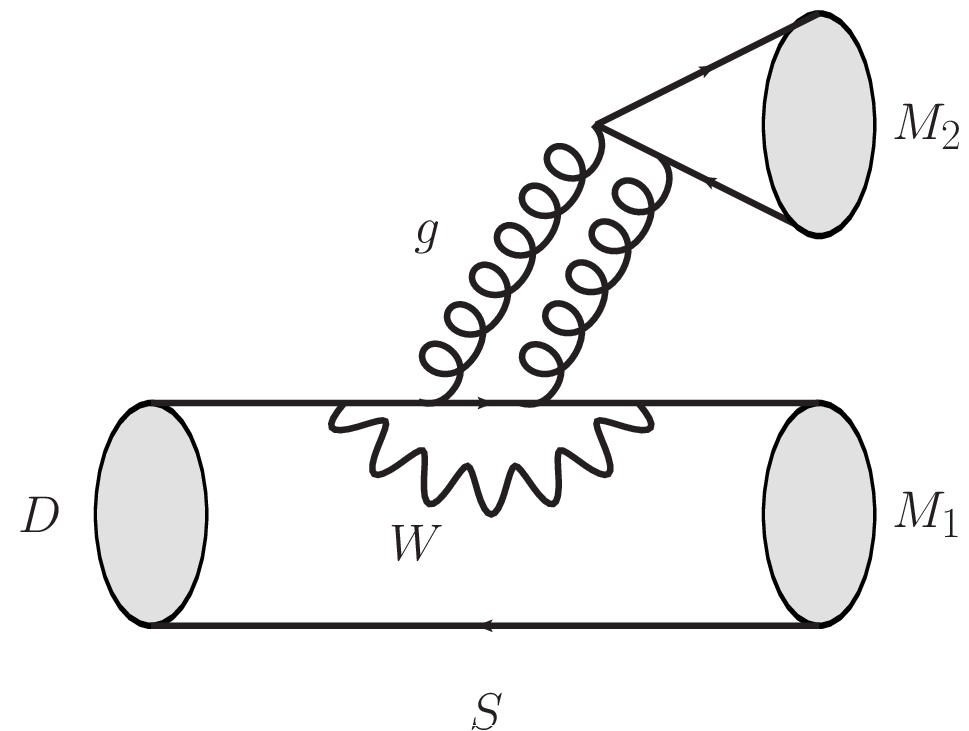}
\includegraphics[width=1.5in]{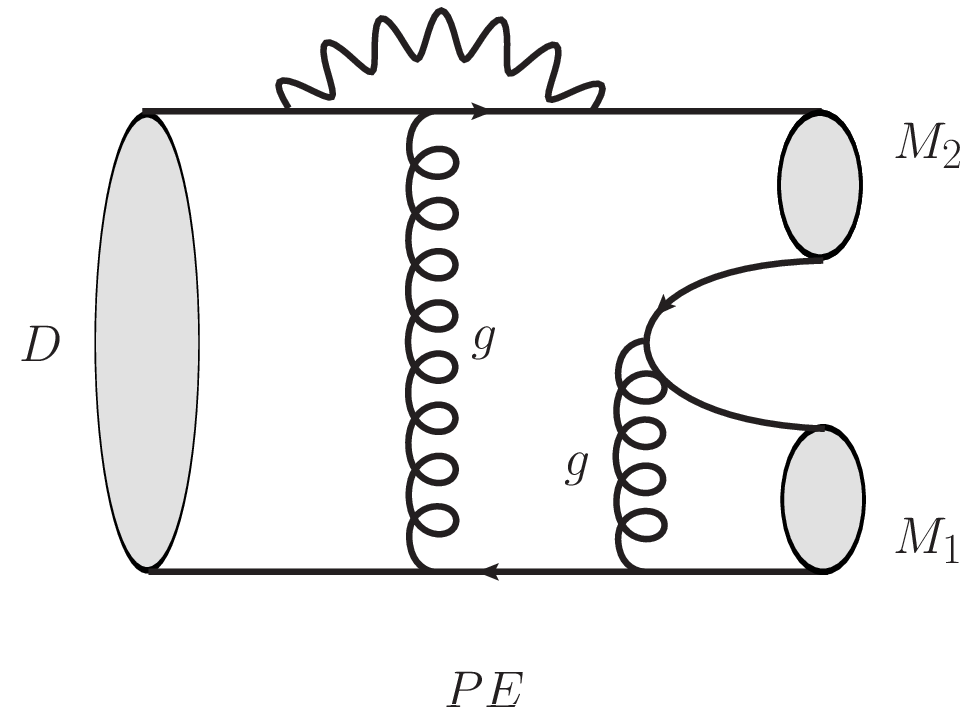}
\includegraphics[width=1.5in]{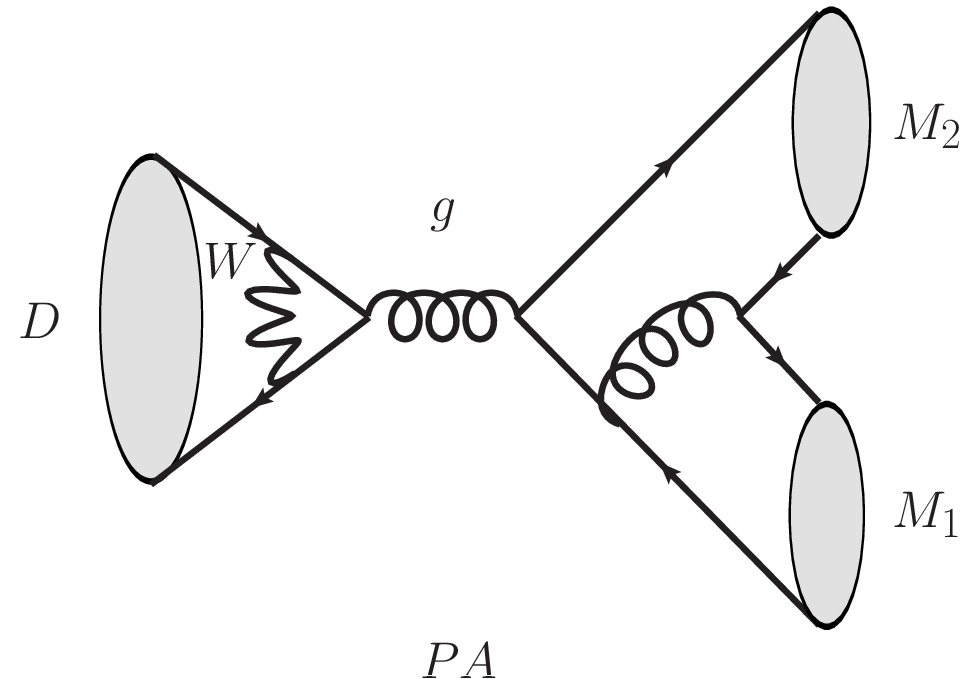}
\vspace*{-1ex}
\caption{ Penguin-type topological diagrams:
(a) QCD-penguin $P$, (b) flavor-singlet QCD-penguin $S$ diagrams with 2 (3) gluon lines for
$M_2$ being a pseudoscalar meson $P$ (a vector meson $V$),  (c) QCD-penguin exchange $P\!E$ and
(d) QCD-penguin annihilation $P\!A$ diagram. In our convention, $M_1$
shares the same spectator quark with the charmed meson, while $M_2$ is an emitted meson. For the $\PA$ diagram with $q\bar q$ produced from the exchanged gluon, $M_1$ is the meson containing the antiquark $\bar q$.
 } \label{Fig:Quarkdiagrams}
\label{Fig1}
\end{figure}
%%%%%%%%%%%%%%%%%%%%%%%%%%%%%%%%%%%

For penguin effects in $D\to M_1M_2$ decays, we consider the QCD-penguin $P$, flavor-singlet QCD-penguin $S$, QCD-penguin exchange $\PE$ and QCD-penguin annihilation $\PA$ as depicted in Fig.~\ref{Fig:Quarkdiagrams} (see~\cite{ChengOh} for details). \footnote{Notice the notation difference (especially $\PE$ and $\PA$) in~\cite{Li:2012} and~\cite{Qin}. The notation of $P_T$, $P_C$, $\PE$ and $\PA$ in~\cite{Li:2012} was changed to $PT$, $PC$, $\PA$ and $\PE$, respectively, in~\cite{Qin}.
In terms of the latter notation,  $P\!T$, $P\!C$, $\PA$ and $\PE$ correspond to our $P$, $S$, $\PE$ and $\PA$, respectively.}
In QCDF, the amplitudes of these QCD-penguin diagrams read~\cite{BN}
\be \label{eq:penguin_general}
\begin{split}
 S^{p}_{M_1 }
   &= {G_F \over \sqrt{2}} [a_3^p(M_1M_2)+\eta\,a_5^p(M_1M_2)]
   \, X^{(D M_1, M_2)}  ~,  \\
 P^{p}_{M_1 }
   &= {G_F \over \sqrt{2}} [a_4^p (M_1M_2)+\eta\,r_\chi^{M_2} a^p_6(M_1M_2)]
   \, X^{(D M_1, M_2)}  ~,   \\
 P\!E^{p}_{M_1 }
 &= {G_F \over \sqrt{2}}
   ~( f_D f_{M_1} f_{M_2})\,{m_V\over m_D\, p_c}\left[ b_3^p \right]_{M_1 M_2}(\epsilon\cdot p_D) ~,
   \\
 P\!A^{p}_{M_1}
 &= {G_F \over \sqrt{2}}
   ~( f_D f_{M_1} f_{M_2}) \,{m_V\over m_D\, p_c}\left[ b_4^p \right]_{M_1 M_2 }(\epsilon\cdot p_D) ~,
\end{split}
\en
where $p=d,s$, $p_c$ is the c.m.~momentum, and $\eta=1$ ($-1$) for $M_1M_2=PV$ ($V\!P$). We have inserted an additional factor of $m_V/(m_D\, p_c)$ in the $\PE$ and $\PA$ amplitudes so that they are expressed in units of $\epsilon\cdot p_D$.
In our convention,
$M_2$ is an emitted meson and $M_1$
shares the same spectator quark with the charmed meson. For the $\PA$ diagram with $q\bar q$ produced from the exchanged gluon, $M_1$ refers to the meson containing the antiquark $\bar q$.
The factorizable matrix element $X$ reads
\be \label{eq:X}
\begin{split}
X^{(DP, V)} &\equiv \langle V| J^{\mu} |0 \rangle
  \langle P| J'_{\mu} |D \rangle
  = 2 f_V \,m_V\,F_1^{DP} (m_{V}^2)(\epsilon\cdot p_D) ~,
\\
X^{(D V,P)} &\equiv \langle P| J^{\mu} |0 \rangle
  \langle V| J'_{\mu} |D \rangle
  = 2 f_P \,m_V\,A_0^{DV} (m_{P}^2)(\epsilon\cdot p_D) ~,
\end{split}
\en
where we have followed the conventional Bauer-Stech-Wirbel definition for form factors~\cite{BSW}.
The chiral factors $r_\chi^{M_2}$ in Eq.~(\ref{eq:penguin_general})  are given by
\begin{eqnarray}
 r_\chi^P(\mu) = {2m_P^2 \over m_c(\mu)(m_2+m_1)(\mu)},  \qquad\quad
 r_\chi^V(\mu) = \frac{2m_V}{m_c(\mu)} ~\frac{f_V^\perp (\mu)}{f_V} ~,
\end{eqnarray}
with $f_V^\perp (\mu)$ being the scale-dependent transverse decay constant of the vector meson $V$.  The flavor operators $a_i^{p}$  have the expressions~\cite{BN}
 \be \label{eq:ai}
  a_i^{p}(M_1M_2) =
 \left(c_i+{c_{i\pm1}\over N_c}\right)N_i(M_2)
  + {c_{i\pm1}\over N_c}\,{C_F\alpha_s\over
 4\pi}\Big[V_i(M_2)+{4\pi^2\over N_c}H_i(M_1M_2)\Big]+{\cal P}_i^{p}(M_2),
 \en
where $C_F=(N_c^2-1)/(2N_c)$ with $N_c$ denoting the number of colors, and contributions from vertex corrections $V_i$, hard spectator interactions $H_i$ and penguin contractions ${\cal P}_i$ have been taken into account~\cite{BN}.  Note that $a_i(M_2M_1)\neq a_i(M_1M_2)$ in general.
The annihilation operators $b_{3,4}^p$ are given by
\be \label{eq:bi}
\begin{split}
b_3^p &= {C_F\over N_c^2}\left[c_3A_1^i+c_5(A_3^i+A_3^f)+N_cc_6 A_3^f\right],
\\
b_4^p &= {C_F\over N_c^2}\left[c_4 A_1^i+c_6 A_2^i\right],
\end{split}
\en
where the annihilation amplitudes $A_{1,2,3}^{i,f}$ are defined in~\cite{BN}.

Flavor amplitude decompositions of the SCS $D\to V\!P$ decays are updated in Table~\ref{tab:CSPV}. Comparing with Table~VI in~\cite{Cheng:2019}, we have made the following improvements: (i) The notation of $\PE_P$ and $\PE_V$  in our previous publications~\cite{Cheng:2012a,Cheng:2019} should be interchanged, i.e. $\PE_P\leftrightarrow \PE_V$. \footnote{ We see from Fig.~\ref{Fig:Quarkdiagrams} that $\PE_P$ ($\PE_V$) is always accompanied by $P_P$ ($P_V$). In our previous publications, we had, for example, $A(D^+\to K^+\ov K^{*0})=\lambda_d A_V+\lambda_s T_V+\lambda_p(P_V^p+\PE_P)$ and
$A(D^+\to \ov K^0 K^{*+})=\lambda_d A_P+\lambda_s T_P+\lambda_p(P_P^p+\PE_V)$ (see Table II of \cite{Cheng:2012a}). Hence, in order to be consistent with the notation in Fig.~\ref{Fig:Quarkdiagrams}, we need to interchange $\PE_P$ and $\PE_V$. This is done in Table~\ref{tab:CSPV}.
}
(ii) Both QCD-penguin annihilation diagrams $\PA_P$ and $\PA_V$ contribute to $D^0\to \pi^\pm \rho^\mp$ and $D^0\to K^\pm K^{*\mp}$ decays. (iii) Singlet QCD-penguin contributions denoted by $S_V$ and $S_P$ for SU(3)-flavor singlets $\eta,\eta',\omega$ and $\phi$ are included. (iv) SU(3) breaking in $W$-exchange amplitudes is taken into account through Eq.~(\ref{eq:SU(3)inW}).  Values of SU(3)-breaking effects are explicitly shown in the following modes:
$D^0\to \pi^0\phi$, $D^+\to \pi^+\phi, K^+\ov K^{*0},\ov K^0K^{*+}$ and $D_s^+\to \pi^+ K^{*0},\pi^0 K^{*+}, K^+\rho^0, K^0\rho^+, K^+\omega$.

%%%%%%%%%%%%%%%%%%%%
\renewcommand{\arraystretch}{1.00}
\begin{table}[tp!]
\caption{Flavor amplitude decompositions for singly Cabibbo-suppressed $D\to V\!P$ decays. Summation over $p=d,~s$ is understood. The symbol $S_V^{\eta_q(\eta_s)}$ implies a singlet amplitude with $M_1=V$ and $M_2=\eta_q(\eta_s)$.
Values of SU(3) breaking effects in  $D^{0}\to \pi^{0}\phi$ and some of $D^+$ and $D_s^+$ decays are shown.
  \label{tab:CSPV}}
\medskip
  {\scriptsize
\begin{ruledtabular}
\begin{tabular}{l l l c  }
 & Mode & Representation   \\
\hline
$D^0$
  & $\pi^+ \rho^-$ & $\lambda_d (T_V+E_P^d)
  + \lambda_p( P_V^p+ \PEV +\PAV +\PAP )$
    \\
  & $\pi^- \rho^+$ & $\lambda_d  (T_P+E_V^d)
  + \lambda_p( P_P^p+ \PEP +\PAV +\PAP )$ \\
  & $\pi^0 \rho^0$ & $\frac12 \lambda_d (-C_P-C_V+E_P^d+E_V^d)$
     $ +\frac12\lambda_p(P_P^p + P_V^p +\PEP +\PEV + \PAP +\PAV )$   \\
  & $K^+ K^{*-}$ & $\lambda_s (T_V+E_P^s) + \lambda_p(P_V^p + \PEV + \PAP+ \PAV)$
   \\
  & $K^- K^{*+}$ & $\lambda_s (T_P+E_V^s) +\lambda_p( P_P^p + \PEP + \PAP+\PAV)$
      \\
  & $K^0 \ol{K}^{*0}$ & $\lambda_d E_V^d+ \lambda_s E_P^s +  \lambda_p(\PAP + \PAV)$
       \\
  & $\ol{K}^0 K^{*0}$ &  $\lambda_d E_P^d+ \lambda_s E_V^s + \lambda_p( \PAP + \PAV)$
       \\
  & $\pi^0 \omega$ & $\frac{1}{2}\lambda_d (-C_V+C_P-E_P^d-E_V^d)
  + \frac{1}{2}\lambda_p(P_P^p + P_V^p + \PEP + \PEV +2S_P)$
     \\
  & $\pi^0 \phi$ & $\frac{1}{\sqrt{2}} \lambda_s (1.23C_P)+\frac{1}{\sqrt{2}} \lambda_p S_P$ \\
  & $\eta \omega$ & $\frac12 \Big[\lambda_d (C_V+C_P+E_V^d+E_P^d)\cos\phi
   - \sqrt{2}\lambda_s C_V\sin\phi$  \\
  & & $\quad +\lambda_p \left[ P_P^p + P_V^p +\PEP + \PEV + \PAP + \PAV+(2S_{\eta_q}+2S_V^{\eta_q}) \cos\phi
  - \sqrt{2}S_V^{\eta_s}\sin\phi \right] \Big]$  \\
  & $\eta\,' \omega$ & $\frac12 \Big[\lambda_d(C_V+C_P+E_V^d+E_P^d)\sin\phi +\sqrt{2}\lambda_s C_V\cos\phi$ \\
  & & $\quad +\lambda_p \left[ P_P^p + P_V^p +\PEP + \PEV + \PAP + \PAV+(2S_{\eta'_q}+2S_V^{\eta'_q}) \sin\phi
  + \sqrt{2}S_V^{\eta'_s}\cos\phi \right] \Big]$  \\
  & $\eta \phi$ & $\lambda_s \left[ {1\over\sqrt{2}}C_P\cos\phi - (E_V^s+E_P^s)\sin\phi \right]
  - \lambda_p(\PAP + \PAV) \sin\phi +{1\over \sqrt{2}}\lambda_p S_{\eta_q}\cos\phi $
   \\
  & $\eta \rho^0$ & ${1\over 2} \Big[ \lambda_d (C_V-C_P-E_V^d-E_P^d)\cos\phi
  - \lambda_s\sqrt{2}C_V\sin\phi \Big]$
     \\
    & & $\quad + {1\over 2}\lambda_p \Big[ (P_P^p + P_V^p + \PEP + \PEV+2S_V^{\eta_q}) \cos\phi
    - \sqrt{2}S_V^{\eta_s} \sin\phi \Big] $  \\
  & $\eta\,' \rho^0$ & ${1\over 2} \Big[ \lambda_d (C_V-C_P-E_V^d-E_P^d)\sin\phi
  + \lambda_s\sqrt{2}C_V\cos\phi \Big]$
     \\
      & & $\quad + {1\over 2}\lambda_p \left[ (P_P^p + P_V^p + \PEP + \PEV +2S_V^{\eta'_q}) \sin\phi + \sqrt{2}S_V^{\eta'_s} \cos\phi \right]$  \\
\hline
$D^+$
  & $\pi^+ \rho^0$ & $\frac{1}{\sqrt{2}} \left[\lambda_d(T_V+C_P-A_P+A_V)
  + \lambda_p(P_V^p - P_P^p + \PEV - \PEP) \right]$ \\
  & $\pi^0 \rho^+$ & $\frac{1}{\sqrt{2}} \left[\lambda_d (T_P+C_V+A_P-A_V)
  + \lambda_p(P_P^p - P_V^p + \PEP - \PEV) \right]$
    \\
  & $\pi^+ \omega$ & $\frac{1}{\sqrt{2}} \left[\lambda_d (T_V+C_P+A_P+A_V)
  + \lambda_p(P_P^p + P_V^p + \PEP + \PEV+ 2S_P) \right]$  \\
  & $\pi^+ \phi$ & $\lambda_s (1.23C_P)+\lambda_p S_P$
   \\
  & $\eta \rho^+$ & ${1\over\sqrt{2}} \left[\lambda_d (T_P+C_V+A_V+A_P)\cos\phi
   - \lambda_s\sqrt{2}C_V\sin\phi \right]$
    \\
  &&$\quad + {1\over\sqrt{2}}\lambda_p \left[(P_P^p + P_V^p + \PEP + \PEV+2S_V^{\eta_q}) \cos\phi
  -\sqrt{2}S_V^{\eta_s}\sin\phi \right]$  \\
  & $\eta\,' \rho^+$ & ${1\over\sqrt{2}} \left[\lambda_d (T_P+C_V+A_V+A_P)\sin\phi
  + \lambda_s\sqrt{2}C_V\cos\phi \right]$
     \\
 &&$\quad + {1\over\sqrt{2}}\lambda_p \left[(P_P^p + P_V^p + \PEP + \PEV+2S_V^{\eta'_q}) \sin\phi
 +\sqrt{2}S_V^{\eta'_s}\cos\phi \right]$ \\
  & $K^+ \ol{K}^{*0}$ & $\lambda_d A_V + \lambda_s (1.22 T_V) +  \lambda_p(P^p_V + \PEV)$
        \\
  & $\ol{K}^0 K^{*+}$ & $\lambda_d A_P + \lambda_s (1.25 T_P) + \lambda_p( P_P^p + \PEP)$
   \\
\hline
$D_s^+$
  & $\pi^+ K^{*0}$ & $\lambda_d (0.74T_V) + \lambda_s A_V + \lambda_p(P_V^p + \PEV)$
   \\
  & $\pi^0 K^{*+}$ & $\frac{1}{\sqrt{2}}
  \left[\lambda_d (0.97C_V) - \lambda_s A_V - \lambda_p(P_V^p + \PEV) \right]$       \\
  & $K^+ \rho^0$ & $\frac{1}{\sqrt{2}}
  \left[\lambda_d (0.77C_P) - \lambda_s A_P - \lambda_p(P_P^p + \PEP)  \right]$ \\
  & $K^0 \rho^+$ & $\lambda_d (0.81T_P) + \lambda_s A_P + \lambda_p(P_P^p + \PEP)$
        \\
  & $\eta K^{*+}$ & ${1\over\sqrt{2}} \left[\lambda_d C_V +  \lambda_s A_V
  + \lambda_p(P_V^p + \PEV +2S_V^{\eta_q} ) \right]\cos\phi$
     \\
  &&$\quad -  \left[ \lambda_s(T_P+C_V+A_P) + \lambda_p(P_P^p + \PEP+ S_V^{\eta_s})  \right]\sin\phi$ \\
  & $\eta\,' K^{*+}$
  & ${1\over\sqrt{2}} \left[\lambda_d C_V + \lambda_s A_V + \lambda_p(P_V^p + \PEV +2S_V^{\eta'_q}) \right] \sin\phi$
     \\
  &&$\quad +  \left[ \lambda_s(T_P+C_V+A_P) + \lambda_p(P_P^p + \PEP +S_V^{\eta'_s}) \right]\cos\phi$ \\
  & $K^+ \omega$ & $\frac{1}{\sqrt{2}}\left[ \lambda_d (0.71C_P) + \lambda_s A_P
  + \lambda_p(P_P^p + \PEP+2S_P) \right]$
   \\
  & $K^+ \phi$ & $\lambda_s(T_V+C_P+A_V) + \lambda_p(P_V^p + \PEV+S_P)$
        \\
\end{tabular}
\end{ruledtabular} }
\end{table}
%
%%%%%%%%%%%%%%%%%%%%%%%%%%%%%%%%%%

Calculations based on QCDF give the hierarchical relation for QCD-penguin amplitudes (see e.g., Table~\ref{tab:KKst} below)
\be
|P_P|\sim |P_V|>|S_V|\sim |\PE_V|\gsim |S_P|\sim |\PE_P|>|\PA_V|\sim |\PA_P|.
\en
%Combining withe the hierarchy pattern Eq.~(\ref{eq:tree_hierarchy}) for tree amplitudes, we have
%\be
% |T_P|\!\! &>&\! |T_V|\gsim|C_{P}|>|C_{V}|\gsim|E_P|>|P_P|\gsim |P_V|\gsim |E_V|  \non \\
%&>&\! |A_{P,V}|
%>|S_V|\sim |\PE_V|\gsim |S_P|\sim |\PE_P|>|\PA_{V,A}|.
%\en
Roughly speaking, $|\PE/P|\sim {\cal O}(10^{-1})$ and $|\PA/P|\sim {\cal O}(10^{-2})$.

%%%%%%%%%%%%%%%%%%%%%%%%%%%%%%%%%%%
\begin{table}[tp!]
\caption{Branching fractions (in units of $10^{-3}$) of singly Cabibbo-suppressed $D\to V\!P$ decays. The predictions made using Solution (S3') have taken into account SU(3) breaking effects under Solution (iv) (see Table~\ref{tab:EPEV}). Experimental branching fractions taken from~\cite{PDG} in conjunction with the narrow width approximation are also quoted in units of $10^{-3}$.
  \label{tab:BFVP}  }
\medskip
\begin{ruledtabular}
\begin{tabular}{l  c c | l c c }
  Mode~~~ & ${\cal B}_{\rm theory}$~~ &  ${\cal B}_{\rm exp}$ & Mode~~~ & ${\cal B}_{\rm theory}$~~ &  ${\cal B}_{\rm exp}$\\
\hline
  $D^0\to\pi^+ \rho^-$ & $5.15\pm0.21$ & $5.15 \pm 0.25$ &  $D^0\to\pi^0 \omega$ & $0.13\pm0.02$   & $0.117\pm0.035$ \\
  $D^0\to\pi^- \rho^+$ & $10.10\pm0.37$  & $10.1 \pm 0.4$ & $D^0\to\pi^0 \phi$ & $0.95\pm0.02$   &  $1.20 \pm 0.04$  \\
  $D^0\to \pi^0 \rho^0$ & $3.79\pm0.12$  & $3.86 \pm 0.23$ & $D^0\to\eta \omega$ & $2.09\pm0.09$ & $1.98\pm0.18$ \\
  $D^0\to K^+ K^{*-}$ & $1.65\pm0.06$  & $1.65 \pm 0.11$ & $D^0\to\eta\,' \omega$ & $0.02\pm0.00$  &---  \\
  $D^0\to K^- K^{*+}$ & $4.56\pm0.15$  & $4.56 \pm 0.21$ & $D^0\to\eta \phi$ & $0.19\pm0.02$ &  $0.181\pm0.034$\footnotemark[3]  \\
  $D^0\to K^0 \ol{K}^{*0}$ & $0.25\pm0.01$  & ~$0.246\pm0.048$~  & $D^0\to\eta \rho^0$ & $0.59\pm0.06$  & ---  \\
  $D^0\to\ol{K}^0 K^{*0}$ & $0.34\pm0.06$  & $0.336\pm0.063$ & $D^0\to\eta\,' \rho^0$ & $0.06\pm0.00$  & ---  \\
\hline
  $D^+\to\pi^+ \rho^0$ & $0.68\pm0.09$ & $0.83 \pm 0.15$ & $D^+\to\eta \rho^+$ & $0.94\pm0.42$  & ---  \\
  $D^+\to\pi^0 \rho^+$ & $4.44\pm0.59$ & --- &  $D^+\to\eta\,' \rho^+$ & $1.23\pm0.11$  & ---  \\
  $D^+\to\pi^+ \omega$ & $0.22\pm0.05$ & $0.28\pm0.06$ & $D^+\to K^+ \ol{K}^{*0}$ & $5.92\pm0.18$  & $3.71\pm0.16$\footnotemark[4]  \\
  $D^+\to\pi^+ \phi$ & $4.87\pm0.10$  & $5.59\pm0.10$ & $D^+\to\ol{K}^0 K^{*+}$ & $16.28\pm0.61$  & $17.6 \pm 1.8$\footnotemark[5]      \\
\hline
  $D_s^+\to\pi^+ K^{*0}$ & $2.06\pm0.08$ & $2.23\pm0.33$\footnotemark[1] & $D_s^+\to\eta K^{*+}$ & $0.46\pm0.19$   & ---  \\
  $D_s^+\to\pi^0 K^{*+}$ & $0.92\pm0.06$  &  $0.75\pm0.25$\footnotemark[2] & $D_s^+\to\eta\,' K^{*+}$ & $0.41\pm0.02$  & ---  \\
  $D_s^+\to K^+ \rho^0$ & $1.22\pm0.06$  & $2.5 \pm 0.4$ &  $D_s^+\to K^+ \omega$ & $0.99\pm0.05$  & $0.87\pm0.25$  \\
  $D_s^+\to K^0 \rho^+$ &  $7.64\pm0.33$  & $5.46\pm0.95$\footnotemark[2] & $D_s^+\to K^+ \phi$ & $0.12\pm0.02$  & $0.18\pm0.04$   \\
\end{tabular}
\footnotetext[1]{The new measurement of $\B(D_s^+\to \pi^+K^{*0})=(2.71\pm0.72\pm0.30)\times 10^{-3}$ from BESIII~\cite{BESIII:DsK0pipi} is taken into account in the world average.}
\footnotetext[2]{Data from BESIII~\cite{BESIII:DsK0pipi}.}
\footnotetext[3]{Apparently, the published result of the BESIII measurement of $D^0\to\eta\phi$~\cite{BESIII:DtophiP} was not taken into account in the world average in the 2020 version of PDG~\cite{PDG}. }
\footnotetext[4]{The new measurement of $\B(D^+\to K^+\ov K^{*0})=(3.10\pm0.46\pm0.68\pm0.18)\times 10^{-3}$ from BESIII~\cite{BESIII:D+toKKst} is taken into account in the world average.}
\footnotetext[5]{The new measurement of $\B(D^+\to K_S K^{*+})=(8.69\pm0.40\pm0.64\pm0.51)\times 10^{-3}$ from BESIII~\cite{BESIII:D+toKKst} is taken into account in the world average.}
\end{ruledtabular}
\end{table}
%
%%%%%%%%%%%%%%%%%%%%%%%%%%%%%%%%%%

\subsection{Branching fractions and SU(3) breaking \label{sec:Branching fractions and SU(3) breaking}}
In the topological-diagram approach, we apply the topological amplitudes extracted from the CF modes to  the SCS decays to predict their rates. It is well known that there exists significant symmetry breaking effects in some of the SCS modes from the flavor SU(3) symmetry limit. In principle, SU(3) breaking effects in the color-allowed $T$ and color-suppressed $C$ tree amplitudes can be estimated provided they are factorizable (in units of $\epsilon\cdot p_D$)
\be
\begin{split}
T_V &= \frac{G_F}{\sqrt2}a_1(\ol K^*\pi)2f_\pi m_{K^*}A_0^{DK^*}(m_\pi^2) ~,
\\
C_P &= \frac{G_F}{\sqrt2}a_2(\ol K^*\pi)2f_{K^*} m_{K^*}F_1^{D\pi}(m_{K^*}^2) ~,
\\
T_P &= \frac{G_F}{\sqrt2}a_1(\ol K\rho)2f_\rho m_\rho F_1^{DK}(m_\rho^2) ~,
\\
C_V &= \frac{G_F}{\sqrt2}a_2(\ol K\rho)2f_K m_\rho A_0^{D\rho}(m_K^2) ~.
\end{split}
\label{factorizationS}
\en
The topological amplitude sizes of $T_V$, $C_P$, $T_P$ and $C_V$ given in Table~\ref{tab:CFVPB} are extracted from a fit to the data of Cabibbo-allowed $D\to V\!P$ decays. They are dominated by $D^0\to K^{*-}\pi^+, \ov K^{*0}\pi^0, K^-\rho^+$ and $\ov K^0\rho^0$, respectively. SU(3) breaking effects in $T_{V,P}$ and $C_{V,P}$ in the SCS modes can be estimated based on factorization. However, as explained in~\cite{Cheng:2019}, we found that the consideration of SU(3) breaking in $T_{V,P}$ and $C_{V,P}$ alone would render even larger deviations from the data. For example, we obtained
$T_V(\pi^+\rho^-) \simeq 0.82\,T_V$, $T_P(\pi^-\rho^+) \simeq 0.92\,T_P$,
$T_V(K^+K^{*-}) \simeq 1.29\,T_V$ and $T_P(K^-K^{*-}) \simeq 1.28\,T_P$. They will lead to too small branching fractions for $D^0\to \pi^+\rho^-$ and $D^0\to \pi^-\rho^+$ and too large  $\B(D^0\to K^+K^{*-})$ and $\B(D^0\to K^-K^{*+})$ compared to experiment.  Hence, the discrepancy becomes even worse.
To circumvent this dilemma, we instead focused on SU(3) breaking in the $W$-exchange amplitudes as we had done in~\cite{Cheng:2019}. By the same token, our predictions on the branching fractions of SCS $V\!P$ decays of $D^+$ and $D_s^+$ in~\cite{Cheng:2019} were made without taking into account SU(3) breaking amplitudes due to the absence of $W$-exchange contributions.

For the experimental branching fractions of the quasi-two-body $D\to V\!P$ decays listed in Table~\ref{tab:BFVP}, we have applied the narrow width approximation (NWA)
\be \label{eq:NWA}
\Gamma(D\to VP\to P_1P_2P)_{\Gamma_V\to 0} =\Gamma(D\to V\!P)\B(V\to P_1P_2)
\en
to extract  $\B(D\to V\!P)$. Notice that this relation is valid only in the narrow width limit, namely, $\Gamma_V\to 0$. Corrections to the NWA have been recently studied in~\cite{Cheng:2020mna,Cheng:2020iwk} for $B$ decays. We shall return to this issue in Sec.~\ref{sec:Finite-width effects}.

Very recently, BESIII~\cite{BESIII:DsK0pipi} has reported the analysis of $D_s^+\to K_S\pi^+\pi^0$ decay and obtained the branching fractions of various modes (see Table~\ref{tab:BFVP}). It turns out that our predictions of $\B(D_s^+\to \pi^+K^{*0})=(3.65\pm0.24) \times 10^{-3}$ and
$\B(D_s^+\to K^0\rho^+)=(11.47\pm0.48) \times 10^{-3}$ %(from here on, all branching fractions are quoted in units of $10^{-3}$, unless otherwise noted)
in~\cite{Cheng:2019} are too large compared to the BESIII new measurements listed in Table \ref{tab:BFVP}. Especially, the latter is larger by a factor of 2. Also, it was already noticed in~\cite{Cheng:2019} that $\B(D_s^+\to K^+\omega)=(2.12\pm0.10) \times 10^{-3}$ was two to three times bigger than the BESIII result of $(0.87\pm0.25) \times 10^{-3}$~\cite{BESIII:DsKomega}. This calls for the necessity of incorporating the SU(3)-breaking effects in the tree amplitudes $T_{V,P}$ and $C_{V,P}$ of the SCS $D_s^+$ decays.

There was a poorly measured branching fraction in the $D^+$ sector, namely, $\B(D^+\to \ov K^0 K^{*+})=(34\pm16)\times 10^{-3}$~\cite{PDG}. Thanks to BESIII, a new measurement of this mode with a significantly improved precision is just available~\cite{BESIII:D+toKKst}. It is based on the amplitude analysis of $D^+\to K^+K_S\pi^0$.
The new world average predominated by BESIII now becomes $\B(D^+\to \ov K^0 K^{*+})=(17.6\pm1.8)\times 10^{-3}$.  Our prediction of $(9.80\pm0.41)\times 10^{-3}$ in \cite{Cheng:2019} is too small by about $4\sigma$.

In light of the discrepancy between theory and several new data of $V\!P$ decays of $D^+$ and $D_s^+$, it calls for SU(3) breaking in SCS decays of both $D^+$ and $D_s^+$.  It appears that we have a rule of thumb in the absence of $W$-exchange: it is necessary to consider its SU(3)-breaking effects if only one of the $T_{V,P}$ and $C_{V,P}$ topological amplitudes appears in the decay amplitude. According to this simple rule, we need to account for SU(3) breaking in the color-allowed or color-suppressed tree topological amplitude only in the following SCS modes:
\be
&& D^0\to \pi^0\phi, \qquad D^+\to \pi^+\phi,~ K^+\ov K^{*0},~K_SK^{*+}, \non \\
&& D_s^+\to \pi^+K^{*0},~\pi^0K^{*+},~K^+\rho^0,~ K^0\rho^+,~K^+\omega.
\en
The SU(3) breaking effects in SCS $D_s\to V\!P$ decays can be estimated from Eq. (\ref{factorizationS}) and factorization
\be \label{eq:SU(3) tree}
\begin{split}
{T_P^{(D_s^+\to K^0\rho^+)}\over T_P}={F_1^{D_sK}(m_\rho^2)\over F_1^{DK}(m_\rho^2)},
&\qquad
{C_V^{(D_s^+\to \pi^0K^{*+})}\over C_V}={f_\pi m_{K^*}\over f_K m_\rho} \,{A_0^{D_sK^*}(m_\pi^2)\over A_0^{D\rho}(m_K^2)}
~, \\
{T_V^{(D_s^+\to \pi^+K^{*0})}\over T_V}={A_0^{D_sK^*}(m_\pi^2)\over A_0^{DK^*}(m_\pi^2)},
&\qquad
{C_P^{(D_s^+\to K^+\rho^0)}\over C_P}={f_\rho m_\rho\over f_{K^*} m_{K^*}} \,{F_1^{D_sK}(m_\rho^2)\over F_1^{D\pi}(m_{K^*}^2)} ~,
\end{split}
\en
and
\be
{C_P^{(D_s^+\to K^+\omega)}\over C_P}={f_\omega m_\omega\over f_{K^*} m_{K^*}} \,{F_1^{D_sK}(m_\omega^2)\over F_1^{D\pi}(m_{K^*}^2)},
\en
where we have assumed that $a_1$ and $a_2$ are process-independent. Likewise, for SCS $D^+\to V\!P$ decays we have
\be \label{eq:SU(3) D+}
\begin{split}
{T_P^{(D^+\to K^+\ov K^{*0})}\over T_P}={f_{K^*} m_{K^*}\over f_{\rho} m_{\rho}}\, {F_1^{D_sK}(m_{K^*}^2)\over F_1^{DK}(m_\rho^2)},
\qquad
{T_V^{(D^+\to \ov K^0K^{*+})}\over T_V}={f_K \over f_\pi}\,{A_0^{DK^*}(m_{K^*}^2)\over A_0^{DK^*}(m_\pi^2)},
\end{split}
\en
and
\be
{C_P^{(D^+\to \pi^+\phi)}\over C_P}={f_\phi m_\phi\over f_{K^*} m_{K^*}} \,{F_1^{D\pi}(m_\phi^2)\over F_1^{D\pi}(m_{K^*}^2)}.
\en

Using the decay constants and form factors together with their $q^2$ dependence evaluated in the covariant confining quark model~\cite{Ivanov:2019nqd}, we show the SU(3) breaking effects in the decay amplitudes in Table~\ref{tab:CSPV}. Now we see from Table~\ref{tab:BFVP} that the agreement with the new BESIII measurements of $D_s^+\to \pi^+K^{*0}$ and $K^0\rho^+$ is  substantially improved. Likewise, the issue with $\B(D_s^+\to K^+\omega)$ as noticed in~\cite{Cheng:2019} is also resolved. However, the predicted $\B(D_s^+\to K^+\rho^0)$ now becomes smaller than experiment. In the topological diagram  approach, $D_s^+\to K^+\rho^0$ and $D_s^+\to K^+\omega$ should have similar rates as their amplitudes are mainly governed by $\lambda_d C_P/\sqrt{2}$. Their SU(3) breaking effects are also similar. It is thus desirable to have an improved measurement of $\B(D_s^+\to K^+\rho^0)$.
For the $D_s^+\to K^+\phi$ mode, if we consider its SU(3) breaking in both $T_V$ and $C_P$ amplitudes, we will have $T_V(K^+\phi)=1.45\,T_V$ and $C_P(K^+\phi)=1.20\,C_P$.  This will lead to $\B(D_s^+\to K^+\phi)=(0.46\pm0.07) \times 10^{-3}$ which is too large compared to the current value of $(0.182\pm0.041) \times 10^{-3}$~\cite{PDG}. This is why we mention in the rule of thumb that if two of the $T_{V,P}$ and $C_{V,P}$ topologies or more appear in the decay amplitude, we should not consider their SU(3)-breaking effects.

In the $D^+$ sector, the predicted $\B(D^+\to \pi^+\phi)=(3.22\pm0.17)\times 10^{-3}$ in the absence of SU(3) breaking is smaller than the experimental value of $(5.59\pm0.10)\times 10^{-3}$ by a factor of 1.7. Thus a SU(3) breaking in $C_P$ is welcome, though it is not large enough in our calculation.
The new prediction of $\B(D^+\to \ov K^0 K^{*+})=(16.3\pm0.6)\times 10^{-3}$ is now in good agreement with the BESIII measurement. Unfortunately,  the original nice agreement between theory and experiment for
$\B(D^+\to K^+\ov K^{*0})$ (see Table~VIII of \cite{Cheng:2019}) is spoiled when the SU(3) breaking in $T_V$ is taken into account. Our result of
$\B(D^+\to K^+\ov K^{*0})=(5.92\pm0.18)\times 10^{-3}$ is too large compared to the measured value of $(3.71\pm0.16)\times 10^{-3}$.
Thus, we face a dilemma that not both $D^+\to\ov K^0K^{*+},K^+\ov K^{*0}$ modes can be accounted for simultaneously in the same SU(3) breaking scheme. For this, we need to wait for further experimental justification.

We have shown in Table~\ref{tab:EPEV} four sets of solutions for SU(3) breaking in the $W$-exchange amplitudes extracted from a fit to eight SCS channels. All the solutions are the same for $E_V^d$ and $E_P^d$ but differ in $E_V^s$ and $E_P^s$. It turns out that
$D^0\to \eta\phi$ is the only channel which receives contributions from both $E_{V,P}^s$: $A(D^0\to \eta\phi)\propto {1\over\sqrt{2}}C_P\cos\phi-(E_V^s+E_P^s)\sin\phi$
(see Table~\ref{tab:CSPV}). Hence, it can be used to discriminate these solutions.
Explicitly, we find (in units of $10^{-3}$)
\be
\B(D^0\to \eta\phi)=0.12\pm0.01,~~ 0.23\pm0.01,~~ 0.008\pm0.005,~~ 0.19\pm0.02
\en
for solutions (i), (ii), (iii) and (iv), respectively. It is obvious that the calculated branching fraction based on solution (iv) agrees better with the measured value of $0.18\pm0.03$ than the other solutions. Therefore, we will stick to solution (iv) for SU(3) breaking in the $W$-exchange amplitudes hereafter.

%%%%%%%%%%%%%%%%%%%%%%%%%%%%%%%%%%%
\begin{table}[tp!]
\caption{Direct \CP asymmetries of singly Cabibbo-suppressed  $D\to V\!P$ decays (in units of $10^{-3}$) in this work, where $a_{\rm dir}^{({\rm tree})}$ denotes \CP asymmetry arising from purely tree amplitudes. The superscript (t+p) denotes tree plus QCD-penguin amplitudes, (t+pe+pa+s) for tree plus $\PE,\PA$ and $S$ amplitudes, (t+pe$^{\rm LD}$) for tree plus long-distance $\PE$ amplitude induced from final-state rescattering and ``tot'' for the total amplitude. As explained in the text, we use solution (iv) for the SU(3) breaking effect in the $W$-exchange amplitudes (see Table~\ref{tab:EPEV}).
The predictions from~\cite{Qin} in the FAT approach with the $\rho-\omega$ mixing are listed in the last column for comparison.
  \label{tab:CPVP}  }
\medskip
  \footnotesize{
\begin{ruledtabular}
\begin{tabular}{ l c  c c r r c}
 Mode
   & $a_{\rm dir}^{({\rm tree})}$ &  $a_{\rm dir}^{({\rm t+p})}$ & $a_{\rm dir}^{({\rm t+pe+pa+s})}$ &  $a_{\rm dir}^{({\rm t+pe^{\rm LD}})}$
     & $a_{\rm dir}^{({\rm tot})}$~~~~ & $a_{\rm dir}^{({\rm tot})}$\cite{Qin} \\
\hline
  $D^0\to\pi^+ \rho^-$  & 0 & $-0.00\pm0.00$ &  $-0.011\pm0.000$& $0.77\pm0.22$ & $0.76\pm0.22$ & $-0.03$ \\
  $D^0\to\pi^- \rho^+$  & 0 &  $0.01\pm0.00$ & $0.008\pm0.001$  & $-0.13\pm0.08$ & $-0.11\pm0.08$ & $-0.01$ \\
   $D^0\to\pi^0 \rho^0$  &0 &  $-0.01\pm0.00$ & $-0.004\pm0.000$  & $0.28\pm0.16$ & $0.27\pm0.16$ & $-0.03$ \\
   $D^0\to K^+ K^{*-}$  & 0 &  $-0.01\pm0.01$ & $0.011\pm0.000$ & $-0.85\pm0.24$ & $-0.85\pm0.24$ & $-0.01$  \\
   $D^0\to K^- K^{*+}$   & 0 &  $-0.03\pm0.00$ & $-0.009\pm0.000$  & $0.08\pm0.09$ & $0.04\pm0.09$ & 0  \\
   $D^0\to K^0 \ol{K}^{*0}$   & $-0.03\pm0.02$ &  $-0.03\pm0.02$ & $-0.03\pm0.02$ & $-0.03\pm0.02$ & $-0.03\pm0.02$ & $-0.7$ \\
   $D^0\to\ol{K}^0 K^{*0}$
     & $1.07\pm0.12$ &  $1.07\pm0.12$ & $1.07\pm0.12$ & $1.07\pm0.12$ & $1.07\pm0.12$ & $-0.7$ \\
   $D^0\to\pi^0 \omega$   & 0 &  $0.04\pm0.00$ & $0.04\pm0.01$ &  $-1.51\pm0.87$ & $-1.43\pm0.87$ & 0.02 \\
   $D^0\to\pi^0 \phi$  & 0 & 0 & $-0.004$~~~  & 0~~~~~~~ & $-0.004$~~~ & $-0.0002$ \\
   $D^0\to\eta \omega$ & $-0.13\pm0.01$ &  $-0.12\pm0.01$ & $-0.13\pm0.01$ & $-0.35\pm0.10$ & $-0.35\pm0.10$ & $-0.1$  \\
   $D^0\to\eta\,' \omega$  & $2.06\pm0.11$ &  $1.93\pm0.11$ & $1.93\pm0.11$ & $1.48\pm0.61$ & $1.23\pm0.60$ & 2.2\\
   $D^0\to\eta \phi$
     & 0 & 0 & $0.009$~~  & 0~~~~~~ & $0.009$~~ & 0.003 \\
   $D^0\to\eta \rho^0$  & $0.45\pm0.03$ & $0.51\pm0.03$ & $0.49\pm0.03$ & $0.16\pm0.30$  & $0.26\pm0.31$ & 1.0 \\
   $D^0\to\eta\,' \rho^0$  & $-0.65\pm0.06$ &  $-0.63\pm0.06$ & $-0.62\pm0.06$ & $-0.17\pm0.23$ & $-0.13\pm0.23$ & $-0.1$ \\
\hline
   $D^+\to\pi^+ \rho^0$  & 0 &  $0.33\pm0.02$ & $0.10\pm0.01$ & $0.83\pm1.36$ & $1.26\pm1.34$ & 0.5\\
   $D^+\to\pi^0 \rho^+$  & 0 &  $0.10\pm0.01$ & $0.04\pm0.00$ & $-0.58\pm0.52$ & $-0.44\pm0.52$ & 0.2 \\
   $D^+\to\pi^+ \omega$  &  0 &  $0.01\pm0.01$ & $0.08\pm0.01$ & $0.93\pm2.28$ & $1.03\pm2.28$ & $-0.05$ \\
   $D^+\to\pi^+ \phi$  & 0 & 0 & $-0.004$~~~ &  0~~~~~~ & $-0.004$~~~ &  $-0.0001$  \\
   $D^+\to\eta \rho^+$ &   $-1.85\pm0.51$ &  $-1.97\pm0.54$ & $-1.93\pm0.55$ & $-2.31\pm0.92$ & $-2.50\pm0.98$ & $-0.6$ \\
   $D^+\to\eta\,' \rho^+$ &  $0.23\pm0.05$ &  $0.20\pm0.05$ & $0.21\pm0.05$ & $0.39\pm0.16$ & $0.34\pm0.15$ & 0.5 \\
   $D^+\to K^+ \ol{K}^{*0}$  &  $-0.11\pm0.01$ &  $-0.14\pm0.01$ & $-0.11\pm0.01$ & $-0.77\pm0.24$ & $-0.80\pm0.24$ & 0.2\\
   $D^+\to\ol{K}^0 K^{*+}$  &  $-0.04\pm0.01$ &  $-0.05\pm0.01$ & $-0.05\pm0.01$ & $-0.06\pm0.06$ & $0.04\pm0.07$ & 0.04 \\
\hline
   $D_s^+\to \pi^+ K^{*0}$ &  $0.18\pm0.02$ &  $0.24\pm0.02$ & $0.19\pm0.02$ & $1.25\pm0.41$ & $1.32\pm0.41$ & $-0.1$  \\
   $D_s^+\to \pi^0 K^{*+}$  &  $0.13\pm0.02$ &   $0.12\pm0.03$ & $0.11\pm0.03$ & $1.35\pm0.40$ & $1.31\pm0.40$ & $-0.2$  \\
   $D_s^+\to K^+ \rho^0$  &  $0.14\pm0.03$ &  $0.11\pm0.02$ & $0.15\pm0.03$ & $-0.26\pm0.12$ & $-0.29\pm0.12$ & 0.3 \\
   $D_s^+\to K^0 \rho^+$  &  $0.06\pm0.02$ &  $0.08\pm0.02$ & $0.08\pm0.02$ & $-0.10\pm0.10$ & $-0.07\pm0.10$ & 0.3 \\
   $D_s^+\to \eta K^{*+}$  &  $1.18\pm0.23$ &  $0.86\pm0.16$ & $0.95\pm0.18$ & $0.95\pm0.75$ & $0.40\pm0.70$ & 1.1 \\
   $D_s^+\to \eta\,' K^{*+}$  &  $-0.19\pm0.04$ &  $-0.16\pm0.04$ & $0.14\pm0.04$ & $-0.33\pm0.19$ & $-0.24\pm0.19$ & $-0.5$ \\
   $D_s^+\to K^+ \omega$  &  $-0.15\pm0.03$ &  $-0.14\pm0.03$ & $-0.16\pm0.03$ & $0.27\pm0.14$ & $0.28\pm0.14$ & $-2.3$ \\
   $D_s^+\to K^+ \phi$  & 0 &  $-0.32\pm0.02$  & $-0.14\pm0.01$ & $-0.88\pm1.61$ & $-1.33\pm1.59$ & $-0.8$ \\
\end{tabular}
\end{ruledtabular}}
\end{table}
%
%%%%%%%%%%%%%%%%%%%%%%%%%%%%%%%%%%

\subsection{Direct \CP violation}
It has been noticed that the QCD-penguin exchange diagrams receive sizable long-distance contributions  from final-state rescattering~\cite{Cheng:2012a}. We shall assume that the long-distance $\PE_V$ and $\PE_P$ are of the same order of magnitude as $E_P$ and $E_V$, respectively. \footnote{As noticed before,
the notation of $\PE_P$ and $\PE_V$  in our previous publications~\cite{Cheng:2012a,Cheng:2019} should be interchanged, $\PE_P\leftrightarrow \PE_V$.
}
For concreteness, we follow~\cite{Cheng:2019} to assign by choice the same magnitude and phase as the $W$-exchange amplitudes with 20\% and $30^\circ$ uncertainties, respectively, so that (in units of $10^{-6}(\epsilon\cdot p_D)$)
\be \label{eq:LDPE}
(P\!E_V)^{\rm LD} \approx (0.58\pm0.12)\,e^{i (283\pm30)^\circ}, \qquad
(P\!E_P)^{\rm LD} \approx (1.69\pm0.34)\,e^{-i (108\pm30)^\circ}.
\en
For simplicity, we shall assume its flavor independence, that is, $(P\!E)^{\rm LD}_d=(P\!E)^{\rm LD}_s$.
\CP asymmetries of the SCS $D\to V\!P$ decays are updated in Table~\ref{tab:CPVP}. The improvements over our previous work (see Table~IX of~\cite{Cheng:2019}) are as follows: (i) In our previous work, the factor of $\epsilon\cdot p_D$ term in the QCD-penguin amplitudes has been replaced by $p_c(m_D/m_V)$, while the tree amplitudes are expressed in terms of $\epsilon\cdot p_D$. This inconsistency is corrected in this work.
(ii) Typos in the code for $D^+\to \pi^0\rho^+$ and $D_s^+\to K^+\phi$ are corrected. The resultant \CP asymmetries  $a^{\rm dir}_{CP}(D^+\to\pi^0\rho^+)=(-0.44\pm0.52)\times 10^{-3}$ and $a^{\rm dir}_{CP}(D_s^+\to K^+\phi)=(-1.33\pm1.59)\times 10^{-3}$ are quite different from the previous values of $(0.08\pm0.11)\times 10^{-3}$ and 0, respectively. (iii) Singlet QCD-penguin contributions as well as weak penguin annihilation ($\PE$ and $\PA$) effects are included under the column denoted by
$a_{\rm dir}^{({\rm t+pe+pa+s})}$. The $D^0\to \pi^0\phi$ and $D^+\to \pi^+\phi$ decays proceed only through the tree diagram $C_P$. Nevertheless, they receive a small flavor-singlet QCD-pengion contribution $S_P$.
Owing to the interference between the tree and singlet QCD-penguin, \CP asymmetries no longer vanish, though they are very small of order $10^{-6}$.

From Table~\ref{tab:CPVP} we identify several golden modes which have large branching fractions and sizeable \CP asymmetries at the order of $10^{-3}$:
\be \label{eq:golden}
D^0\to \pi^+\rho^-, K^+K^{*-},\qquad D^+\to \eta\rho^+, K^+\overline{K}^{*0},  \qquad
D_s^+\to \pi^+ K^{*0}, \pi^0K^{*+}.
\en
It is interesting to notice that the \CP asymmetry difference defined by
\be \label{eq:acp VP}
\Delta a_{CP}^{V\!P}\equiv a_{CP}(K^+K^{*-})-a_{CP}(\pi^+\rho^-),
\en
in analogy to $\Delta A_{C\!P}$ defined in Eq.~(\ref{eq:LHCb:2019}) for the corresponding $P\!P$ final states, is predicted to be $(-1.61\pm0.33)\times 10^{-3}$, which is very similar to the recently observed \CP asymmetry difference between $D^0\to K^+K^-$ and $D^0\to\pi^+\pi^-$. This is an attractive and measurable observable in the near future.
It is thus desirable to first search for \CP violation in the aforementioned golden modes.

Since the quasi-two-body decays $D^0\to \pi^+\rho^-$ ($D^0\to \pi^-\rho^+$) and $D^0\to K^+K^{*-}$ ($D^0\to K^-K^{*+}$) are connected by an interchange of all $d$ and $s$ quarks, there exists a general $U$-spin relation between corresponding \CP-rate differences~\cite{Gronau:2000zy}
\be
|A(D\to f)|^2-|A(\ov D\to \bar f)|^2=-\left(|A(UD\to Uf)|^2-|A(U\ov D\to U\bar f)|^2\right)
~,
\en
where $U$ denotes a $U$-spin transformation ($d \leftrightarrow s$) and the overall minus sign on the right-hand side comes from a change in the CKM factors.
Hence,
\be
\begin{split}
\Gamma(D^0\to \pi^+\rho^-)-\Gamma(\ov D^0\to \pi^-\rho^+)
&= -\left(\Gamma(D^0\to K^+K^{*-})-\Gamma(\ov D^0\to K^-K^{*+})\right),
\\
\Gamma(D^0\to \pi^-\rho^+)-\Gamma(\ov D^0\to \pi^+\rho^-)
&= -\left(\Gamma(D^0\to K^-K^{*+})-\Gamma(\ov D^0\to K^+K^{*-})\right).
\end{split}
\en
Since the $C\!P$-averaged rates of $D^0\to \pi^\pm\rho^\mp$ and $D^0\to K^\pm K^{*\mp}$ are identical in the $U$-spin limit, we are led to the $U$-spin relations
\be \label{eq:U-spin relations}
\begin{split}
a^{\rm dir}_{CP}(D^0\to K^+K^{*-})
&= -a^{\rm dir}_{CP}(D^0\to \pi^+\rho^-),
\\
a^{\rm dir}_{CP}(D^0\to K^-K^{*+})
&= -a^{\rm dir}_{CP}(D^0\to \pi^-\rho^+).
\end{split}
\en
\CP asymmetries for the above-mentioned two modes have equal magnitudes and opposite signs. It is evident from Table~\ref{tab:CPVP} that the first $U$-spin relation in Eq.~\eqref{eq:U-spin relations} is approximately respected in our calculation, but not so for the second $U$-spin relation.

To see the effect of long-distance QCD-penguin exchange through final-state rescattering, we compare \CP violation under the two columns $a_{\rm dir}^{(t+pe^{\rm LD})}$ and $a_{\rm dir}^{({\rm tot})}$ in Table~\ref{tab:CPVP}. It is obvious that \CP asymmetries of the following modes are dominated by the long-distance QCD-penguin exchange:
\be
&& D^0\to \pi^+\rho^-,~ \pi^-\rho^+,~\pi^0\rho^0, ~K^+K^{*-}, ~ K^-K^{*+},~\pi^0\omega, ~\eta\,\omega,  \non \\
&& D^+\to\pi^0\rho^+, ~\pi^+\omega,~ \eta\rho^+,~ K^+\ov K^{*0},  \\
&& D_s^+\to \pi^+K^{*0}, ~\pi^0 K^{*+}, ~K^+\rho^0, ~K^0\rho^+,~K^+\omega. \non
\en
The six golden modes listed in Eq.~(\ref{eq:golden}) all belong to this category.
 To see this, we consider the four decay modes $D^0\to \pi^\pm\rho^\mp$ and $D^0\to K^\pm K^{*\mp}$.  From Table~\ref{tab:CSPV}, their amplitudes read
\be \label{eq:Amp pirho KKst}
\M(D^0\to \pi^+\rho^-) &=& \lambda_d(T_\rho+E_\pi^d)+\lambda_dP_\rho^d+ \lambda_s P_\rho^s -\lambda_b(\PE_\rho+\PA_\rho+\PA_\pi), \non \\
\M(D^0\to \rho^+\pi^-) &=& \lambda_d(T_\pi+E_\rho^d)+\lambda_dP_\pi^d+ \lambda_s P_\pi^s -\lambda_b(\PE_\pi+\PA_\rho+\PA_\pi),  \\
\M(D^0\to K^+K^{*-}) &=& \lambda_s(T_{K^*}+E^s_K)+\lambda_dP_{K^*}^d+ \lambda_s P_{K^*}^s -\lambda_b(\PE_{K^*}+\PA_{K^*}+\PA_K),  \non \\
\M(D^0\to K^{*+}K^-) &=& \lambda_s(T_K+E^s_{K^*})+\lambda_dP_K^d+ \lambda_s P_K^s -\lambda_b(\PE_K+\PA_{K^*}+\PA_K), \non
\en
with $\lambda_p\equiv V_{cp}^*V_{up}~(p=d,s,b)$. Direct \CP asymmetries are expressed by
\be \label{eq:DCPV pirho KKst}
a_{\CP}^{\rm dir}(\pi^+\rho^-) &=& 1.30\times 10^{-3} \left| {P_V^s+P\!E_V^s+P\!A_V^s+\PA_P^d \over T_V+E_P^d+\Delta P}\right|_{_{\pi\rho}}\sin\theta_{_{\pi\rho}}, \non \\
a_{\CP}^{\rm dir}(\rho^+\pi^-) &=& 1.30\times 10^{-3} \left| {P_P^s+P\!E_P^s+P\!A_P^s+\PA_V^d \over T_P+E_V^d+\Delta P}\right|_{_{\rho\pi}}\sin\theta_{_{\rho\pi}}, \non \\
a_{\CP}^{\rm dir}(K^+K^{*-}) &=& -1.30\times 10^{-3} \left| {P_V^d+P\!E_V^d+P\!A_V^s+\PA_P^d \over T_V+E_P^s-\Delta P}\right|_{_{K\!K^*}}\sin\theta_{_{K\!K^*}},  \\
a_{\CP}^{\rm dir}(K^{*+}K^-) &=& -1.30\times 10^{-3} \left| {P_P^d+P\!E_P^d+P\!A_P^s+\PA_V^d \over T_P+E_V^s-\Delta P}\right|_{_{K^*\!K}}\sin\theta_{_{K^*\!K}}, \non
\en
where $\theta_{_{\pi\rho}}$ is the phase of $(P_V^s+\cdots+\PA_P^d)$ relative to $(T_V+E_P^d+\Delta P)$ and likewise for the other phases.

%%%%%%%%%%%%%%%%%%%%%%%%%%%%%%%%%%%
\begin{table}[t]
\caption{Numerical results of the topological tree and penguin amplitudes of $D^0\to \pi^\pm\rho^\mp$ and $D^0\to K^\pm K^{*\mp}$ in units of $10^{-6}(\epsilon\cdot p_D)$.
Only central values are presented here.
  \label{tab:KKst}}
  \medskip
  \scriptsize{
\begin{ruledtabular}
\begin{tabular}{ l c c c c c c}
$D^0\to \pi^+\rho^-$ & $T_\rho$ & $E_\pi^d$ & $P_\rho^d$ & $P_\rho^s$ & $\PE_\rho$ & $\PA_\rho+\PA_\pi$ \\
  & 2.18 & $-0.02+1.03i$ & $(0.92+0.48i)\,10^{-1}$ & $(0.96+0.44i)\, 10^{-1}$ & $4.60\times 10^{-2}$ & $4.96\times 10^{-4}$ \\
 \hline
$D^0\to \pi^-\rho^+$  & $T_\pi$ & $E_\rho^d$ & $P_\pi^d$ & $P_\pi^s$ & $\PE_\pi$ & $\PA_\rho+\PA_\pi$ \\
  & $1.72+3.11i$ & $0.00-0.20i$ & $-(0.74+0.66i)\,10^{-1}$ & $-(0.78+0.64i)\,10^{-1}$ & $-2.51\times 10^{-2}$ &  $4.96\times 10^{-4}$ \\
  \hline
$D^0\to K^+K^{*-}$ & $T_{K^*}$ & $E_K^s$ & $P_{K^*}^d$ & $P_{K^*}^s$ & $\PE_{K^*}$ & $\PA_{K^*}+\PA_K$  \\
  & 2.18 & $-0.08+0.67i$ & $(1.18+0.61i)\,10^{-1}$ & $(1.22+0.56i)\,10^{-1}$ & $6.06\times 10^{-2}$ & $1.78\times 10^{-3}$ \\
 \hline
 $D^0\to K^-K^{*+}$  & $T_K$ & $E_{K^*}^s$ & $P_K^d$ & $P_K^s$ & $\PE_K$ & $\PA_{K^*}+\PA_K$ \\
  & $1.72+3.11i$ & $-0.83+0.44i$ & $-(1.12+1.04i)\,10^{-1}$ & $-(1.18+1.00i)\,10^{-1}$ & $-2.77\times 10^{-2}$ &  $1.78\times 10^{-3}$ \\
\end{tabular}
\end{ruledtabular}
}
\end{table}
%%%%%%%%%%%%%%%%%%%%%%%%%%%%%%%%%

In Table~\ref{tab:KKst} we show the numerical values of the topological tree and penguin amplitudes of $D^0\to \pi^\pm\rho^\mp$ and $D^0\to K^\pm K^{*\mp}$. The magnitudes of $T_{V,P}$ and $E_{V,P}^{d,s}$ are taken from Tables~\ref{tab:CFVPB} and \ref{tab:EPEV}, while QCD-penguins $P,\PE$ and $\PA$ are calculated in QCDF through Eq.~(\ref{eq:penguin_general}).
It follows that
\be \label{eq:P/T pirho}
\left({P_V^s\over T_V}\right)_{\pi^+\rho^-}=
0.048\,e^{i24.7^\circ}, \qquad
\left({P_V^s+\PE_V^s+\PA_V^d+\PA_P^s\over T_V+E_P^d+\Delta P}\right)_{\pi^+\rho^-}=
0.062\,e^{-i8.9^\circ},
\en
for $D^0\to \pi^+\rho^-$ and
\be  \label{eq:P/T KKst}
\left({P_V^d\over T_V}\right)_{K^+K^{*-}}=
0.061\,e^{i27.3^\circ}, \qquad
\left({P_V^d+\PE_V^d+\PA_V^s+\PA_P^d\over T_V+E_P^s-\Delta P}\right)_{K^+K^{*-}}=
0.086\,e^{i1.2^\circ},
\en
for $D^0\to K^+K^{*-}$. It is clear that the magnitude of $P/T$, which is of order 0.20 in the $P\!P$ sector (see Eq.~(\ref{eq:PoverT QCDF})), is further suppressed in the $V\!P$ sector.
Substituting Eqs.~(\ref{eq:P/T pirho}) and (\ref{eq:P/T KKst}) into
Eq.~(\ref{eq:DCPV pirho KKst}) yields
\be \label{eq:acp pi+rho-}
a_{\rm dir}^{(t+p+pe+pa)}(\pi^+\rho^-)=-1.2\times 10^{-5}, \qquad
a_{\rm dir}^{(t+p+pe+pa)}(K^+ K^{*-})=-2.4\times 10^{-6},
\en
which are consistent with the results shown in Table~\ref{tab:CPVP}.
\footnote{In the absence of tree contributions to \CP violation (i.e. $a_{\rm dir}^t=0$),
$a_{\rm dir}^{(t+p+pe+pa)}$ will be the sum of $a_{\rm dir}^{(t+p)}$ and $a_{\rm dir}^{(t+pe+pa)}$ shown in Table~\ref{tab:CPVP}. Therefore, $a_{\rm dir}^{(t+p+pe+pa)}(\pi^+\rho^-)=(-1.2\pm0.6)\times 10^{-5}$ and $a_{\rm dir}^{(t+p+pe+pa)}(K^+K^{*-})=(-2.2\pm0.4)\times 10^{-6}$.}
Obviously,  the phase of $(P_V^d+\PE_V^d+\PA_V^d+\PA_P^s)$ is almost compensated by that of $(T_V+E_P^s-\Delta P)$, that is, $\theta_{_{K^+K^{*-}}}=1.2^\circ$.
so that the resulting \CP asymmetry in $D^0\to K^+K^{*-}$ is vanishing small of order $10^{-6}$. Likewise, $\theta_{_{\pi^+\rho}}=-8.9^\circ$ also implies a very small direct \CP violation in $D^0\to \pi^+\rho^-$, of order $10^{-5}$. For completeness, we give the results for $\pi^-\rho^+$ and $K^-K^{*+}$ modes
\be \label{eq:acp pi-rho+}
a_{\rm dir}^{(t+p+pe+pa)}(\pi^-\rho^+)=2.1\times 10^{-5}, \qquad
a_{\rm dir}^{(t+p+pe+pa)}(K^- K^{*+})=-3.8\times 10^{-5}.
\en

After including the long-distance contributions to $\PE$ via final-state rescattering (see Eq.~(\ref{eq:LDPE})) we have
\be
\begin{split}
\left({P_V^s+\PE_V^s+\PA_V^d+\PA_P^s+\PE_V^{\rm LD}\over T_V+E_P^d+\Delta P}\right)_{\pi^+\rho^-} &=&
0.704\,e^{i77^\circ}
~,
\\
\left({P_V^d+\PE_V^d+\PA_V^s+\PA_P^d+\PE_V^{\rm LD}\over T_V+E_P^s-\Delta P}\right)_{K^+K^{*-}} &=& 0.771\,e^{i84^\circ}
~.
\end{split}
\en
It follows that
\be  \label{eq:aCPKKst}
a_{\CP}^{\rm dir}(\pi^+\rho^-)=0.89\times 10^{-3}
~, \qquad
a_{\CP}^{\rm dir}(K^+K^{*-})=-1.00\times 10^{-3}
~,
\en
consistent with the predictions of  $a_{\CP}^{\rm dir}(\pi^+\rho^-)=(0.76\pm0.22)\times 10^{-3}$ and $a_{\CP}^{\rm dir}(K^+K^{*-})=(-0.85\pm0.24)\times 10^{-3}$ shown in Table~\ref{tab:CPVP}. \footnote{As stated in~\cite{Cheng:2019}, predictions in Tables~\ref{tab:BFVP} and~\ref{tab:CPVP} are made by sampling $10^4$ points in the parameter space, assuming that each of the parameters has a Gaussian distribution with the corresponding central value and symmetrized standard deviation.  Then the predicted values are the mean and standard deviation of data computed using the $10^4$ points. This explains the difference between Eq.~(\ref{eq:aCPKKst}) and the central values presented in Table~\ref{tab:CPVP}.
}

Two remarks are in order: (i) In the topological amplitude approach, the magnitude and the relative strong phase of each individual topological tree amplitude in charm decays can be extracted from the data. Consequently, direct \CP asymmetries in charmed meson decays induced at the tree level can be reliably estimated.
Since the contribution from $\PA$ to $D^0\to K^0\ov K^{*0}$ and  $D^0\to \ov K^0 K^{*0}$ are very small compared to the $W$-exchange amplitudes, \CP asymmetries in these two modes are induced at the tree level, in analogue to $D^0\to K_SK_S$. We find a large \CP asymmetry at the per mille level for $D^0\to K_SK^{*0}$ but not for $D^0\to K_S\ov K^{*0}$:
\be
a_{\CP}^{\rm dir}(D^0\to K_SK^{*0}) =(1.07\pm0.12)\times 10^{-3}
~.
\en
This is consistent with the upper limit of $0.3\%$ found in \cite{Nierste}.
In view of the current efforts in search of \CP asymmetry in $D^0\to K_SK_S$,
it is worthwhile to pursue that in the $D^0\to K_SK^{*0}$ decay as well.
(ii) Although the magnitude of $\PE$ and $\PA$ is smaller than $P$ (see e.g., Table~\ref{tab:KKst}), we see from Table~\ref{tab:CPVP} that $a_{\rm dir}^{(t+pe+pa)}$ is not necessarily smaller than $a_{\rm dir}^{(t+p)}$ in some channels. It depends on the relative phases between $T$ and $P$ as well as between $T$ and $\PE$.

\subsection{Comparison with the FAT approach}
From Table~\ref{tab:CPVP} it is evident that the predicted \CP asymmetries given in~\cite{Qin} based on the FAT approach are generally smaller than ours by one to two orders of magnitude.
To see the underlying reason, we consider QCD-penguin exchange $\PE$ and annihilation $\PA$ amplitudes. It follows from Eqs.~(\ref{eq:penguin_general}) and (\ref{eq:bi}) that
\be \label{}
\PE &=& {G_F\over\sqrt{2}} (f_D f_P f_V){C_F\over N_c^2}\left[c_3A_1^i+c_5 A_3^i+(c_5+N_cc_6) A_3^f\right], \non \\
\PA &=& {G_F\over\sqrt{2}} (f_D f_P f_V){C_F\over N_c^2}\left[c_4 A_1^i+c_6 A_2^i\right],
\en
where the superscripts `$i$' and `$f$' refer to gluon emission from the initial and final-state quarks, respectively. The subscript `$k$' on $A_k^{i,f}$ refer to one of the three possible Dirac structures: $k=1$ for $(V-A)\otimes(V-A)$, $k=2$ for $(V-A)\otimes(V+A)$ and $k=3$ for $-2(S-P)\otimes(S+P)$~\cite{BN}. The amplitudes $A_k$ have the expressions
\be
A_1 &\propto& \la M_1M_2|(\bar uq)_{_{V-A}}\otimes(\bar qc)_{_{V-A}}|D\ra,  \non \\
A_2 &\propto &  \la M_1M_2|(\bar uq)_{_{V-A}}\otimes(\bar qc)_{_{V+A}}|D\ra,   \\
A_3 &\propto & -2\la M_1M_2|(\bar uq)_{_{S+P}}\otimes(\bar qc)_{_{S-P}}|D\ra,   \non
\en
with $(\bar q_1q_2)_{_{S\pm P}}\equiv \bar q_1(1\pm\gamma_5)q_2$ and $(\bar q_1q_2)_{_{V\pm A}}\equiv \bar q_1\gamma_\mu(1\pm\gamma_5)q_2$. Now $A_k^f$ corresponds to the factorizable contribution,
while $A_k^i$ to the nonfactorizable contribution of $A_k$. It turns out $A_1^f=A_2^f=0$ because of helicity suppression, but not so for $A_3^f$ owing to its $(S-P)(S+P)$ structure for the four-quark operator; that is,
\be
A_3^f\propto -2\la M_1M_2|(\bar uq)_{_{S+P}}|0\ra\la 0|(\bar qc)_{_{S-P}}|D\ra \ ,
\en

In~\cite{Qin} the factorizable $\PE_{V(P)}^f$ amplitude has the expression
\be
\PE_{V(P)}^f={G_F\over\sqrt{2}}a_6(\mu)(-2)\la V\!P(PV)|(\bar uq)_{_{S+P}}|0\ra\la 0|(\bar qc)_{_{S-P}}|D\ra.
\en
It was evaluated in the pole model by assuming its dominance by resonant pseudoscalars. Explicitly, it reads
\footnote{For the $PP$ case, it was assumed in~\cite{Li:2012} that $\la P_1P_2|(\bar q_1q_2|0\ra$ was dominated by the lowest scalar resonance, so that $\la P_1P_2|(\bar q_1q_2|0\ra=\la P_1P_2|S\ra\la S|\bar q_1q_2|0\ra=g_ST_S^{\rm BW}(q^2)m_S\bar f_S$ with $T_S^{\rm BW}$ being the Breit-Wigner line shape for the scalar resonance $S$. However, the light scalars such as $\sigma(500)$, $f_0(980)$ and $a_0(980)$ were not taken into account in~\cite{Li:2012}.  In QCDF, $A_3^f$ is expressed in terms of the twist-2 LCDA $\Phi_M$ and the twist-3 one $\Phi_m$.
}
\be
\la V\!P(PV)|(\bar uq)_{_{S+P}}|0\ra\to \la  V\!P(PV)|H_s|P^*\ra {1\over m_D^2-m_{P^*}^2} \la P^*|(\bar uq)_{_{S+P}}|0\ra,
\en
where $P^*$ represents the pole resonant pseudoscalar meson and $H_s$ is the corresponding strong Hamiltonian. Since $\la PV|H_s|P^*\ra =-\la  V\!P|H_s|P^*\ra$~\cite{Qin}, this leads to the relation $A_3^f(PV)+A_3^f(V\!P)=0$. Take $D^0\to \pi^+\rho^-$ as an example where $P_V$ and $\PE_V^f$ are given by~\cite{Qin}
\be
\begin{split}
P_V
&= {G_F\over\sqrt{2}} \left[ a_4(\mu)-r_\chi^\pi a_6(\mu) \right] 2f_\pi m_\rho A_0^{D\rho}(m_\pi^2)
~, \\
\PE_V^f
&= {G_F\over \sqrt{2}}a_6(\mu)g_{PPV}r_\chi^\pi f_{P^*}f_D m_D^2{1\over m_D^2-m^2_{P^*}}
~.
\end{split}
\en
It was claimed in~\cite{Qin} that there was a numerical coincidence that $P_V$ and $\PE_V^f$ canceled each other. As a consequence, \CP asymmetries in $D^0\to \pi^\pm\rho^\mp$ and $D^0\to K^\pm K^{*\mp}$ decays are very small of order $10^{-5}$.

In QCDF, $A_3^f$ is expressed in terms of the twist-2 light-cone distribution amplitude (LCDA) $\Phi_M$ and the twist-3 one $\Phi_m$. A direct evaluation of the weak annihilation diagram with the four-quark operator $(\bar uq)_{_{S+P}}\otimes(\bar qc)_{_{S-P}}$ yields~\cite{BN}
\be \label{eq:A3f}
\begin{split}
A_3^f(V\!P)
&= \pi\alpha_s\int^1_0 dxdy\left\{r_\chi^V\Phi_P(x)\Phi_{v}(y){2(1+\bar x)\over \bar x^2 y}-r_\chi^P\Phi_V(y)\Phi_p(x){2(1+y)\over \bar x y^2}\right\}
~, \\
A_3^f(PV)
&= \pi\alpha_s\int^1_0 dxdy\left\{r_\chi^P\Phi_V(x)\Phi_{p}(y){2(1+\bar x)\over \bar x^2 y}+r_\chi^V\Phi_P(y)\Phi_v(x){2(1+y)\over \bar x y^2}\right\}
~,
\end{split}
\en
with $\bar x=1-x$ and $\bar y=1-y$. It is obvious that $A_3^f(PV)\neq -A_3^f(V\!P)$ in the QCDF approach, contrary to the aforementioned pole model assumption. It is known that the integrals in Eq.~(\ref{eq:A3f}) involve endpoint divergences. We shall follow~\cite{BBNS99} to model the endpoint divergence $X\equiv\int^1_0 dx/\bar x$ in the penguin annihilation  diagram as
\be \label{eq:XA}
 X_A = \ln\left({m_D\over \Lambda_h}\right)(1+\rho_A e^{i\phi_A}),
\en
with $\Lambda_h$ being a typical hadronic scale of 0.5 GeV. The expressions of $A_3^f$ can be further simplified by using the asymptotic distribution amplitudes $\Phi_{P,V}(x)=6x\bar x$, $\Phi_p(x)=1$ and $\Phi_v(x)=3(x-\bar x)$~\cite{BN}:
\be
\begin{split}
A_3^f(V\!P) &\approx  6\pi\alpha_s\left[3r_\chi^V(2X_A-1)(2-X_A)-r_\chi^P \Big(2X_A^2-X_A\Big)\right]
~, \\
A_3^f(PV) &\approx  6\pi\alpha_s\left[-3r_\chi^P(2X_A-1)(2-X_A)+r_\chi^V \Big(2X_A^2-X_A\Big)\right]
~.
\end{split}
\en
In principle, one can also add the superscripts `$V\!P$' and `$PV$' to distinguish penguin annihilation effects in $D\to V\!P$ and $D\to PV$ decays. Unfortunately, unlike $B$ decays we do not have any
knowledge about the parameters $\rho_A$ and $\phi_A$ for charm decays. Therefore, we shall consider the ``default" value $X_A=\ln (m_D/\Lambda_h)$ in this work. The numerical values of $\PE$ and $\PA$ shown in Table~\ref{tab:KKst} are obtained using the default value of $X_A$.
We see from this table that $-\lambda_b\PE_V$ is smaller than $\lambda_d P_d+\lambda_s P_s$ and they are of the same sign. Therefore, there is no cancellation between $P_V$ and $\PE_V$ in QCDF!

In short, before the consideration of long-distance contribution to $\PE$ through final-state rescattering, the predicted \CP asymmetries in $D^0\to \pi^\pm\rho^\mp$ and $D^0\to K^\pm K^{*\mp}$ in QCDF are also very small, of order $10^{-5}$ ($K^+K^{*-}$ is further suppressed; see Eqs. (\ref{eq:acp pi+rho-}) and (\ref{eq:acp pi-rho+})). However, they are very small in QCDF for a reason quite different from FAT: the phase angles $\theta_{_{\pi\rho}}$, $\theta_{_{\rho\pi}}$, $\theta_{_{KK^*}}$ and $\theta_{_{K^*K}}$ in Eq. (\ref{eq:DCPV pirho KKst}) become smaller or even close to zero after including the contributions from $\PE,\PA$ and $W$-exchange to the ratio of $P/T$.
It is the long-distance QCD-penguin exchange that explains why our predictions of \CP asymmetries in  $D^0\to \pi^\pm\rho^\mp$ and $D^0\to K^\pm K^{*\mp}$ are much bigger than those in the FAT approach.

%%%%%%%%%%%%%%%%%%%%%%%%%%%%%%%%%%%%%%%%%%%%%%%%%%
\section{\CP violation in three-body $D$ decays through vector resonances \label{sec:3-body DCPV}}
%%%%%%%%%%%%%%%%%%%%%%%%%%%%%%%%%%%%%%%%%%%%%%%%%%

In this section we would like to study the Dalitz plots of \CP asymmetry distributions in
some of the SCS $D\to P_1P_2P_3$ decays. In particular, we shall focus on the three-body $D^0\to K^+K^-\pi^0$, $D^0\to \pi^+\pi^-\pi^0$, $D^+\to K^+K_S\pi^0$ and $D_s^+\to K^0\pi^+\pi^0$ decays as they receive contributions from quasi-two-body $D^0\to K^\pm K^{*\mp}$, $D^0\to \pi^\pm\rho^\mp,\pi^0\rho^0$, $D^+\to K_SK^{*+},K^+\ov K^{*0}$ and $D_s^+\to \pi^{+,0}K^{*{0,+}}$ decays, respectively. Due to the interference between various vector resonances carrying the same mass but different electric charges, the magnitude and sign of local \CP asymmetry vary from region to region. Large \CP asymmetries can occur in some localized regions of phase space.
For previous studies of \CP violation in three-body $D$ decays, see~\cite{Zhou:2018suj,Dery:2021mll}.  For a study of three-body hadronic $D$ decay amplitudes within the framework of QCDF, the reader is referred to \cite{Boito:2017jav}. For the proper three-body approaches using more complex theoretical background
in Fadeev techniques, Khuri-Treiman and Triangle singularities, see \cite{Niecknig:2015ija,Nakamura:2015qga,Magalhaes:2011sh,Aoude:2018zty}. 

In principle, \CP asymmetry may also arise from the interference between $P$- and $S$-wave contributions. For example, a significant
\CP asymmetry in the $\rho(770)$ region coming from the interference between the $\rho(770)$ and $S$-wave resonances has been observed by the LHCb \cite{LHCb:pipipi,LHCb:pippippim}.  See also the discussions in \cite{Boito:2017jav}.  Owing to the lack of knowledge on scalar resonances, we will not pursue this direction in this work.

\subsection{$D^0\to K^+K^-\pi^0$}
Consider the Dalitz plot analysis of  $D^0\to K^+K^-\pi^0$ in the overlapped region of the vector resonances $K^{*+}(892)$ and $K^{*-}(892)$ so that
\be \label{eq:DtoKKpi}
A(D^0\to K^+(p_1)K^-(p_2)\pi^0(p_3))\approx A_{K^{*+}}+A_{K^{*-}},
\en
with
\be
A_{K^{*\pm}} \equiv A(D^0\to K^\mp K^{*\pm}\to K^+K^-\pi^0).
\en
The explicit expression of $A_{K^{*+}}$ reads
\be  \label{eq:AKst}
\begin{split}
A_{K^{*+}}
&= \la K^+(p_1)\pi^0(p_3)|{\cal L}|K^{*+}\ra T_{K^*}^{\rm BW}(s_{13})\la K^-(p_2)K^{*+}|H_W|D^0\ra
\\
&= g^{K^{*+}\to K^+\pi^0}F(s_{13},m_{K^*}) \epsilon^*\cdot(p_3-p_1)T_{K^*}^{\rm BW}(s_{13})\M(D^0\to K^-K^{*+})
~,
\end{split}
\en
where $\M(D^0\to K^-K^{*+})$ has been given in Eq.~\eqref{eq:Amp pirho KKst} and we have considered the relativistic Breit-Wigner line shape for $K^*(892)$,
\be
T_{K^*}^{\rm BW}(s)={1\over s-m_{K^*}^2+im_{K^*}\Gamma_{K^*}(s)}
~,
\en
and a mass-dependent width
\be \label{eq:GammaR}
\Gamma_{K^*}(s)=\Gamma_{K^*}^0\left( {q\over q_0}\right)^3
{m_{K^*}\over \sqrt{s}} {X^2_1(q)\over X^2_1(q_0)},
\en
with $q=|\vec{p}_1|=|\vec{p}_3|$ being the c.m.~momentum in the rest frame of $K^{*+}$, $q_0$ the value of $q$ when $s_{13}$ is equal to $m_{K^*}^2$, and $X_1$  a Blatt-Weisskopf barrier factor given by
\be \label{eq:XJ}
X_1(z)=\sqrt{1\over (z\,r_{\rm BW})^2+1},
\en
with $r_{{\rm BW}}\approx 4.0\,{\rm GeV}^{-1}$. In Eq.~(\ref{eq:GammaR}), $\Gamma_{K^*}^0$ is the nominal total width of $K^*$ with $\Gamma_{K^*}^0=\Gamma_{K^*}(m_{K^*}^2)$. Likewise,
the amplitude $A_{K^{*-}}$ is given by
\be
A_{K^{*-}} = g^{K^{*-}\to K^-\pi^0}F(s_{23},m_{K^*}) \epsilon^*\cdot(p_2-p_3)
T_{K^*}^{\rm BW}(s_{23})\M(D^0\to K^-K^{*+}).
\en

To derive Eq.~(\ref{eq:AKst}), we have taken the Lagrangian
\be \label{eq:Lag KstKpi}
\begin{split}
{\cal L} =&
- ig^{K^*\to K\pi}\Big(
{1\over\sqrt{2}}K^{*+\mu}K^- \stackrel{\leftrightarrow}{\partial}_\mu \pi^0
+ {1\over\sqrt{2}}K^{*-\mu}\pi^0 \stackrel{\leftrightarrow}{\partial}_\mu K^+
- {1\over\sqrt{2}}\ov K^{*0\mu}\pi^0 \stackrel{\leftrightarrow}{\partial}_\mu K^0
\\
& \qquad\qquad\qquad  +K^{*+\mu}\ov K^0 \stackrel{\leftrightarrow}{\partial}_\mu \pi^-
- {1\over\sqrt{2}}K^{*0\mu}\ov K^0 \stackrel{\leftrightarrow} {\partial}_\mu \pi^0
\Big)
\end{split}
\en
to obtain $\la K^+\pi^0|K^{*+\mu}K^- \stackrel{\leftrightarrow}{\partial}_\mu \pi^0 |K^{*+}\ra = i\epsilon^*\cdot(p_3-p_1)$.
The form factor $F(s,m_{K^*})$ in the amplitudes $A_{K^{*\pm}}$ is introduced  for the following reason. The coupling $|g^{K^{*\pm}\to K^\pm\pi^0}|=3.15$ is extracted from the measured $K^*(892)$ width through the relation
\be \label{}
 \Gamma_{K^{*\pm}\to K^\pm\pi^0}={q_0^3\over 6\pi m_{K^*}^2}g_{K^{*\pm}\to K^\pm\pi^0}^2.
\en
When $K^*(892)$ is off the mass shell, especially when $s_{13}$ is approaching the upper bound of $(m_D-m_K)^2$, it is necessary to account for the off-shell effect. For this purpose, we shall follow~\cite{Cheng:2020iwk} to introduce a form factor $F(s,m_R)$ parameterized as
\be \label{eq:FF for coupling}
F(s,m_R)=\left( {\Lambda^2+m_R^2 \over \Lambda^2+s}\right)^n,
\en
with the cutoff $\Lambda$ not far from the resonance,
\be
\Lambda=m_R+\beta\Lambda_{\rm QCD},
\en
where the parameter $\beta$ is expected to be of order unity. We shall use $n=1$, $\Lambda_{\rm QCD}=250$~MeV and $\beta=1.0\pm0.2$ in subsequent calculations.

%====================================================================
\begin{figure}[t]
\begin{center}
\centering
 {\includegraphics[scale=0.59]{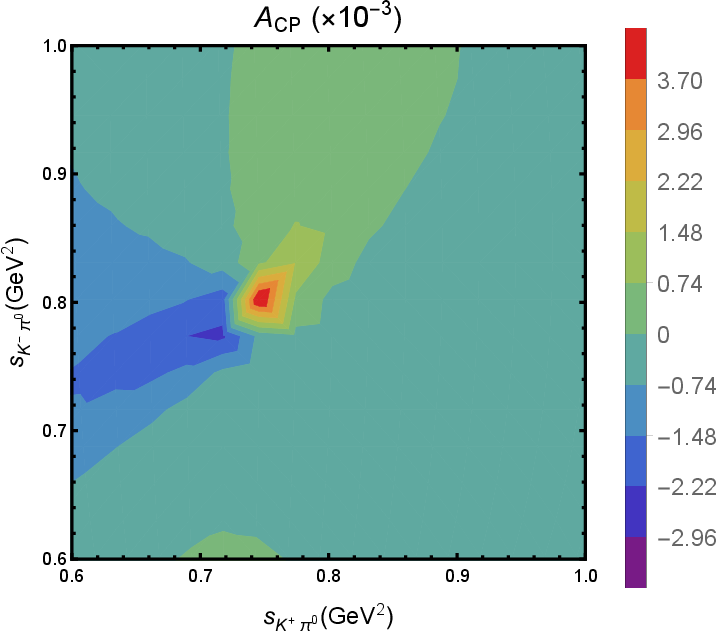}
}
\vspace{0.1cm}
\caption{\CP asymmetry distribution of $D^0\to K^+K^-\pi^0$ in the overlapped regions of $K^*(892)^+$ and $K^*(892)^-$.
}
\label{fig:KKst_CP}
\end{center}
\end{figure}
%=====================================================================

The sum over the polarizations of $K^*$ yields the familiar relation
\be
\sum_\lambda \epsilon^{*\mu}(k,\lambda)\epsilon^\nu(k,\lambda)=-g^{\mu\nu}+{k^\mu k^\nu\over m^2_{K^*}}
~.
\en
However, when the transversality condition $\epsilon^\mu k_\mu=0$ is imposed, the mass in the denominator of the second term should be replaced by the invariant mass $m_{13}=\sqrt{s_{13}}$ or $m_{23}=\sqrt{s_{23}}$~\cite{Asner:2003gh}, where $s_{13}\equiv (p_1+p_3)^2= s_{K^+\pi^0}$ and $s_{23}\equiv (p_2+p_3)^2= s_{K^-\pi^0}$. Replacing $\epsilon\cdot p_D$ in $A_{K^{*+}}$ by ${1\over 2}\epsilon\cdot (p_D+p_2)$ and $\epsilon\cdot p_D$ in $A_{K^{*-}}$ by ${1\over 2}\epsilon\cdot (p_D+p_1)$, it is straightforward to show that
\be \label{eq:AKst_explicit}
\begin{split}
A_{K^{*+}} =& -{1\over 2}\left(s_{23}-s_{12}+{(m_D^2-m_K^2)(m_K^2-m_\pi^2)\over s_{13}}\right) g^{K^{*+}\to K^+\pi^0}
\\
& \qquad\times F(s_{13},m_{K^*})T_{K^*}^{\rm BW}(s_{13})\tilde\M(D^0\to K^+K^{*-})
~, \\
A_{K^{*-}} =& {1\over 2}\left(s_{13}-s_{12}+{(m_D^2-m_K^2)(m_K^2-m_\pi^2)\over s_{23}}\right) g^{K^{*-}\to K^-\pi^0}
\\
& \qquad\times F(s_{23},m_{K^*})T_{K^*}^{\rm BW}(s_{23})\tilde\M(D^0\to K^-K^{*+})
~,
\end{split}
\en
where $\M=\tilde\M(\epsilon\cdot p_D)$.
Notice that the term in the big parentheses can be recast to
\be
s_{23}-s_{12}+{(m_D^2-m_K^2)(m_K^2-m_\pi^2)\over s_{13}}=4\vec{p}_1\cdot\vec{p_2}=4|\vec{p}_1||\vec{p}_2|\cos\theta,
\en
where $|\vec{p}_1|$ and $|\vec{p}_2|$ are the momenta of $K^+$ and $K^-$, respectively, in the rest frame of $K^+$ and $\pi^0$. Hence, the angular distribution is proportional to the Legendre polynomial $P_1(\cos\theta)$.

From Eqs.~(\ref{eq:AKst_explicit}) and (\ref{eq:Amp pirho KKst}), we obtain
\be
\begin{split}
\B(D^0\to K^-K^{*+}\to K^+K^-\pi^0) &= 1.39\times 10^{-3}
~, \\
\B(D^0\to K^+K^{*-}\to K^+K^-\pi^0) &= 0.51\times 10^{-3}
~,
\end{split}
\en
and
\be
\begin{split}
a_{C\!P}(D^0\to K^-K^{*+}\to K^+K^-\pi^0) &= 0.06\times 10^{-3}
~, \\
a_{C\!P}(D^0\to K^+K^{*-}\to K^+K^-\pi^0) &= -0.99\times 10^{-3}
~.
\end{split}
\en
Note that the branching fractions given above are consistent with that in Table \ref{tab:BFVP}
in conjunction with the narrow width approximation. Likewise, the calculated \CP asymmetries
are consistent with Table \ref{tab:CPVP}.

We show in Fig.~\ref{fig:KKst_CP} the Dalitz plot of the \CP asymmetry distribution in the overlapped regions of $K^*(892)^+$ and $K^*(892)^-$. Owing to the interference between
$D^0\to K^+K^{*-}\to K^+K^-\pi^0$ and $D^0\to K^-K^{*+}\to K^+K^-\pi^0$,
local \CP violation varies from region to region. For example, $a_{C\!P}$ is of the percent level in the region around $(s_{K^+\pi^0},s_{K^-\pi^0})=(0.75,0.80)\,{\rm GeV}^2$ (the pink area in  Fig.~\ref{fig:KKst_CP}) and it becomes negative
of order $-2.0\times 10^{-3}$ in the vicinity of  $(s_{K^+\pi^0},s_{K^-\pi^0})=(0.70,0.77)\,{\rm GeV}^2$.

Note that an additional relative phase $\delta$ is introduced to Eq.~(\ref{eq:DtoKKpi}) in~\cite{Zhou:2018suj} so that
\be
A(D^0\to K^+(p_1)K^-(p_2)\pi^0(p_3))=A_{K^{*+}}+e^{i\delta}A_{K^{*-}}
=|A_{K^{*+}}|e^{i\delta_{K^{*+}}}+e^{i\delta}|A_{K^{*-}}|e^{i\delta_{K^{*-}}}.
\en
Experimentally, the phase $\delta_{\rm exp}$ defined by
\be
A(D^0\to K^+(p_1)K^-(p_2)\pi^0(p_3))
=(|A_{K^{*+}}|+e^{i\delta_{\rm exp}}|A_{K^{*-}}|)e^{i\delta_{K^{*+}}}
\en
was measured to be $-35.5^\circ\pm1.9^\circ\pm2.2^\circ$ by BaBar~\cite{BaBar:KpKmpi0}. Neglecting the small penguin contributions to $D^0\to K^\pm K^{*\mp}$, we have
\be
\delta=\delta_{\rm exp}+\delta_{K^{*+}}-\delta_{K^{*-}}.
\en
We find $\delta_{K^{*+}}=8.82^\circ$ and $\delta_{K^{*-}}=-29.74^\circ$ from Eq.~(\ref{eq:Amp pirho KKst}) and Table~\ref{tab:KKst}. Our $\delta=(3.06\pm2.91)^\circ$ is to be compared with $\delta=(-51.85\pm2.91)^\circ$ obtained in~\cite{Zhou:2018suj}.

\subsection{$D^0\to \pi^+\pi^-\pi^0$}
The Dalitz analysis of the three-body $D^0\to \pi^+\pi^-\pi^0$ decay by BaBar~\cite{BaBar:D0pippimpi0} shows fit fractions of $(68.1\pm0.6)\%$, $(34.6\pm0.8)\%$ and $(25.9\pm1.1)\%$ from $\rho(770)^+\pi^-$, $\rho(770)^-\pi^+$ and $\rho(770)^0\pi^0$, respectively. Therefore, the quasi-two-body $D^0\to \pi\rho(770)$ decay gives the dominant contributions to $D^0\to \pi^+\pi^-\pi^0$. Contributions from $\rho(1450)\pi$ and $\rho(1700)\pi$ are small.
Scalar meson contributions are further suppressed.

Consider the Dalitz plot of $D^0\to \pi^+\pi^-\pi^0$ in the overlapped region of the vector resonances $\rho(770)^\pm$ and $\rho(770)^0$ so that
\be
A(D^0\to \pi^+(p_1)\pi^-(p_2)\pi^0(p_3))=A_{\rho^+}+A_{\rho^-}+A_{\rho^0},
\en
with
\be \label{eq:A_rho}
\begin{split}
A_{\rho^+}
&\equiv A(D^0\to \rho^+\pi^-\to \pi^+\pi^-\pi^0)
\\
&= g^{\rho\to \rho\pi} \epsilon^*\cdot(p_1-p_3)F(s_{13},m_\rho) T_\rho(s_{13})\M(D^0\to \rho^+\pi^-)
~, \\
A_{\rho^-}
&\equiv A(D^0\to \rho^-\pi^+\to \pi^+\pi^-\pi^0)
\\
&= g^{\rho\to \rho\pi} \epsilon^*\cdot(p_3-p_2)F(s_{23},m_\rho) T_\rho(s_{23})\M(D^0\to \rho^+\pi^-)
~, \\
A_{\rho^0}
&\equiv A(D^0\to \rho^0\pi^0\to \pi^+\pi^-\pi^0)
\\
&= g^{\rho\to \rho\pi} \epsilon^*\cdot(p_2-p_1)F(s_{12},m_\rho) T_\rho(s_{12})\M(D^0\to \rho^0\pi^0)
~,
\end{split}
\en
where we have used the $\rho\pi\pi$ interaction Lagrangian
\be
{\cal L}=-ig^{\rho\to \pi\pi} \left(
   \rho^{+\mu}\pi^0 \stackrel{\leftrightarrow}{\partial}_\mu \pi^-
+  \rho^{-\mu}\pi^+ \stackrel{\leftrightarrow}{\partial}_\mu \pi^0
+  \rho^{0\mu}\pi^- \stackrel{\leftrightarrow}{\partial}_\mu \pi^+
\right).
\en
Numerically, the on-shell coupling is $|g^{\rho\to \pi\pi}|=6.0$.
Hence,
\be \label{eq:Arho_2}
\begin{split}
A_{\rho^+} &= -{1\over 2}(s_{12}-s_{23}) g^{\rho\to \pi\pi}F(s_{13},m_\rho) T_\rho(s_{13})\tilde\M(D^0\to \rho^+\pi^-)
~, \\
A_{\rho^-} &= {1\over 2}(s_{12}-s_{13}) g^{\rho\to \pi\pi}F(s_{23},m_\rho)T_\rho(s_{23})\tilde\M(D^0\to \rho^-\pi^+)
~, \\
A_{\rho^0} &= {1\over 2}(s_{13}-s_{23}) g^{\rho\to \pi\pi}F(s_{12},m_\rho)T_\rho(s_{12})\tilde\M(D^0\to \rho^0\pi^0)
~. \end{split}
\en

%====================================================================
\begin{figure}[t]
\begin{center}
\centering
 {
 \includegraphics[scale=0.59]{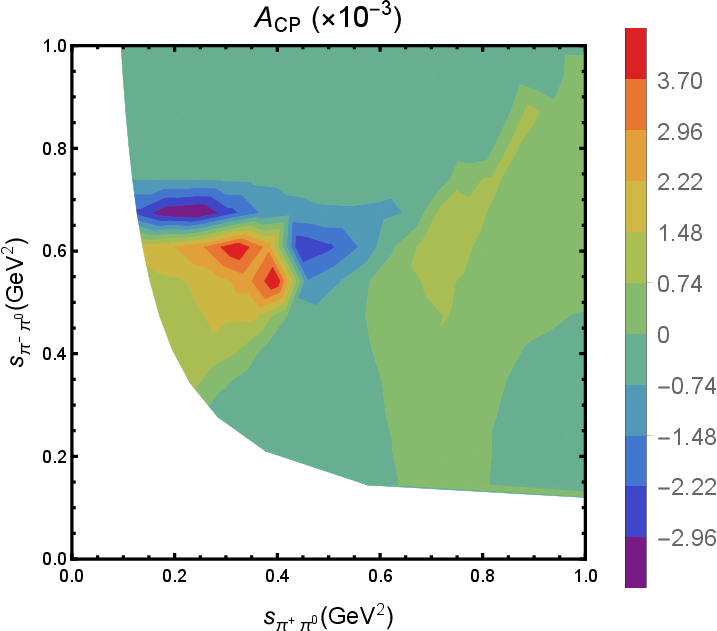}\quad
 \includegraphics[scale=0.59]{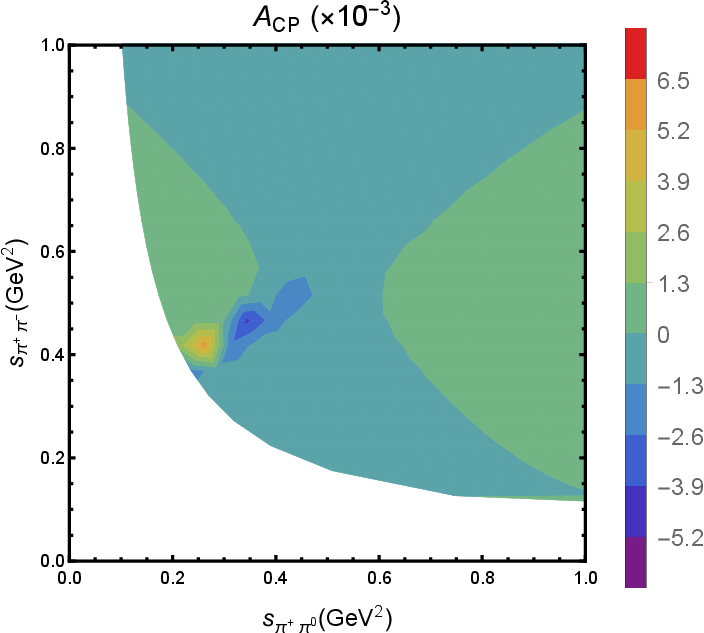}
}
\vspace{0.1cm}
\caption{\CP asymmetry distribution of $D^0\to \pi^+\pi^-\pi^0$ in the overlapped region of $\rho(770)^\pm$ and $\rho(770)^0$ with $s_{\pi^+\pi^0}$ versus $s_{\pi^-\pi^0}$ ($s_{\pi^+\pi^-}$) in the left (right) panel.}
\label{fig:3pi_CP}
\end{center}
\end{figure}
%=====================================================================

Since the $\rho(770)$ resonance is broad, a popular choice for describing its line shape is the Gounaris-Sakurai (GS) model~\cite{Gounaris:1968mw} given by
\be
T_\rho^{\rm GS}(s)={ 1+D \, \Gamma_\rho^0/m_\rho \over s-m^2_{\rho}-f(s)+im_{\rho}\Gamma_{\rho}(s)}
~.
\label{eq: T GS}
\en
In this model, the real part of the pion-pion scattering amplitude with an intermediate $\rho$ exchange calculated from the dispersion relation is taken into account by the $f(s)$ term in the propagator of $T_\rho^{\rm GS}(s)$. Unitarity far from the pole mass is thus ensured.
Explicitly,
\be \label{eq:f(s)}
f(s)=\Gamma_\rho^0{m_\rho^2\over q_0^3}\left[ q^2[h(\sqrt{s})-h(m_\rho)]+(m_\rho^2-s)q_0^2\left.{dh\over ds}\right\vert_{m_\rho}\right]
~,
\en
and
\be
h(s)={2\over \pi}{q\over \sqrt{s}}\ln\left( {\sqrt{s}+2q\over 2m_\pi}\right)
~, \qquad
\left.{dh\over ds}\right\vert_{m_\rho}=h(m_\rho)\left[ {1\over 8q_0^2}-{1\over 2m_\rho^2}\right]+{1\over 2\pi m_\rho^2}
~.
\en
The constant parameter $D$ is given by
\be
D
= {3\over \pi}\,{m_\pi^2\over q_0^2}\ln \left( {m_\rho+2q_0\over 2m_\pi} \right)
+ {m_\rho\over 2\pi q_0}-{m_\pi^2 m_\rho\over \pi q^3_0}
~.
\en

We find
\be
\begin{split}
\B(D^0\to \pi^+\rho^-\to \pi^+\pi^-\pi^0) &= 5.59\times 10^{-3}
~, \\
\B(D^0\to \pi^-\rho^+\to \pi^+\pi^-\pi^0) &= 13.11\times 10^{-3}
~, \\
\B(D^0\to \pi^0\rho^0\to \pi^+\pi^-\pi^0) &= 5.07\times 10^{-3}
~,
\end{split}
\en
and
\be
\begin{split}
a_{C\!P}(D^0\to \pi^+\rho^-\to \pi^+\pi^-\pi^0) &= 0.886\times 10^{-3}
~, \\
a_{C\!P}(D^0\to \pi^-\rho^+\to \pi^+\pi^-\pi^0) &= -0.150\times 10^{-3}
~, \\
a_{C\!P}(D^0\to \pi^0\rho^0\to \pi^+\pi^-\pi^0) &= 0.295\times 10^{-3}
~.
\end{split}
\en

\CP asymmetry distributions in the Dalitz plot are illustrated in Fig.~\ref{fig:3pi_CP}. The interference between the vector resonances $\rho^\pm$ and $\rho^0$ yields rich information on the local \CP asymmetries in the Dalitz plot. It can reach the $4\times 10^{-3}$ level in the regions of $(s_{\pi^+\pi^0},s_{\pi^-\pi^0})=(0.31,0.60)\,{\rm GeV}^2$ and $(0.38,0.54)\,{\rm GeV}^2$ and  becomes very negative
of order $-5\times 10^{-3}$ at  $(s_{\pi^+\pi^0},s_{\pi^-\pi^0})=(0.25,0.67)\,{\rm GeV}^2$.

\subsection{$D^+\to K^+K_S\pi^0$}
For the SCS $D^+\to V\!P$ decays, we see from Table~\ref{tab:CPVP} that the following quasi-two-body decays $\pi^+\rho^0$, $\pi^0\rho^+$, $\eta\rho^+$ and $K^+\ov K^{*0}$ are predicted to have large \CP asymmetries at the per mille level, of order $(1.0\sim 2.5)\times 10^{-3}$.
The first two modes $\pi^+\rho^0$ and $\pi^0\rho^+$ mainfest in the decays
$D^+\to \pi^+\pi^+\pi^-$ and $\pi^+\pi^0\pi^0$, respectively. Due to the presence of two neutral pions, it is na{\"i}vely expected that the former 3-body decay has a rate larger than the latter. Experimentally, it is the other way around: $\B(D^+\to \pi^+\pi^+\pi^-)=(3.27\pm0.18)\times 10^{-3}<  \B(D^+\to \pi^+\pi^0\pi^0)=(4.7\pm0.4)\times 10^{-3}$~\cite{PDG}. The $\pi^+\pi^+\pi^-$ mode is dominated by the $S$-wave contribution to $\pi^+\pi^-$, while the fit fraction of $\pi^+\rho^0$ is about 25\%, yielding $\B(D^+\to \pi^+\rho^0)=(0.83\pm0.15)\times 10^{-3}$~\cite{PDG}.
An amplitude analysis of $D^+\to \pi^+\pi^0\pi^0$ is still not available. Our prediction of $\B(D^+\to \pi^0\rho^+)=(4.44\pm0.59)\times 10^{-3}$ (see Table~\ref{tab:BFVP}) implies that this mode is dominated by $\pi^0\rho^+$.

The $D^+\to\eta\pi^+\pi^0$ decay receives resonant contributions from $\rho^+$ and $a_0(980)^\pm$. To study the $\rho^+$ contribution experimentally, the contribution from $a_0(980)$ can be eliminated with a simple cut by demanding $M(\pi^+\pi^0)<1$ GeV.\
\footnote{It is known that ${a}_0(980)$ has a width of $50-100$~MeV~\cite{PDG} though.}
This will allow us to extract the branching fraction of $D^+\to\eta\rho^+$, which is predicted to be $(0.94\pm0.42)\times 10^{-3}$ in this work (see Table~\ref{tab:BFVP}).
The branching fraction of $D^+\to\eta\pi^+\pi^0$ was measured by CLEO to be $(1.38\pm0.35)\times 10^{-3}$. Improved measurement by BESIII~\cite{BESIII:Detapipi} recently yields $(2.23\pm0.15\pm0.10)\times 10^{-3}$.

The $D^+\to K^+K^-\pi^+$ mode is dominated by the quasi-two-body decays $\pi^+\phi$, $K^+\ov K^{*0}(892)$ and $K^+\ov K^{*0}(1430)$ with the fit fractions of order 28\%, 26\% and 19\%, respectively~\cite{Rubin:2008aa}. In this work we find \CP asymmetries with the values of $-3.7\times 10^{-6}$ and $(-0.80\pm0.24)\times 10^{-3}$, respectively, for $\pi^+\phi$ and $K^+\ov K^{*0}(892)$. In~\cite{Li:2019hho} it  was claimed that a sizable \CP violation could occur in $D^+\to K^+\ov K^{*0}(1430)$ within the FAT approach.  A search of \CP violation in $D^+\to K^+K^-\pi^+$ has been carried out by LHCb without any evidence~\cite{LHCb:DKKpi}.

%====================================================================
\begin{figure}[t]
\begin{center}
\centering
 {\includegraphics[scale=0.59]{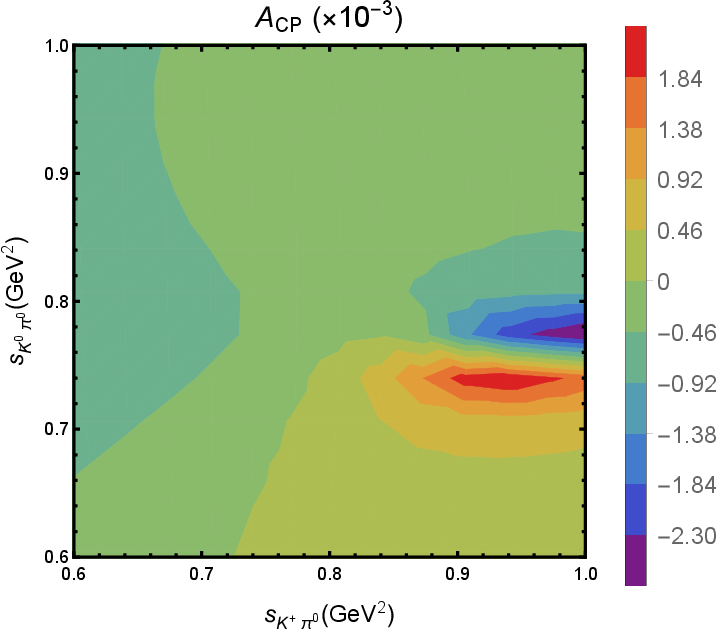}
}
\vspace{0.1cm}
\caption{\CP asymmetry distribution of $D^+\to K^+\ov K^0\pi^0$ in the overlapped regions of $K^*(892)^+$ and $\ov K^*(892)^0$.
}
\label{fig:DpKKpi_CP}
\end{center}
\end{figure}
%=====================================================================

In short, among the aforementioned four $D^+\to V\!P$ modes, analysis of $\pi^0\rho^+$ and $\eta\rho^+$ is not available yet and the branching fraction of $\pi^+\rho^0$ is too small.
Finally, we  turn to the decay $D^+\to K^+\ov K^0\pi^0$ whose amplitude analysis was recently performed by BESIII \cite{BESIII:D+toKKst}. It is dominated by the quasi-two-body decays $K_SK^{*+}$ and $K^+\ov K^{*0}$  with the fit fractions of order 57\% and 10\%, respectively,
\be \label{eq:DtoKKpi}
A(D^+\to K^+(p_1)\ov K^0(p_2)\pi^0(p_3))\approx A_{K^{*+}}+A_{\bar K^{*0}},
\en
with
\be
A_{K^{*+}} \equiv A(D^+\to K^{*+}\ov K^0\to K^+\ov K^0\pi^0), \qquad
A_{\bar K^{*0}} \equiv A(D^+\to K^{+}\ov K^{*0}\to K^+\ov K^0\pi^0).
\en
Following the similar analysis for $D^0\to K^+K^-\pi^0$, we obtain
\be \label{eq:AKst_explicit}
A_{K^{*+}} &=& -{1\over 2}\left(s_{23}-s_{12}+{(m_D^2-m_K^2)(m_K^2-m_\pi^2)\over s_{13}}\right) g^{K^{*+}\to K^+\pi^0} \non \\
&& \times F(s_{13},m_{K^*})T_{K^*}^{\rm BW}(s_{13})\tilde\M(D^+\to \ov K^0K^{*+}),  \\
A_{\bar K^{*0}} &=& -{1\over 2}\left(s_{13}-s_{12}+{(m_D^2-m_K^2)(m_K^2-m_\pi^2)\over s_{23}}\right) g^{\bar K^{*0}\to \ov K^0\pi^0} \non \\
&& \times F(s_{23},m_{K^*})T_{K^*}^{\rm BW}(s_{23})\tilde\M(D^+\to K^+\ov K^{*0}), \non
\en
where
\be
\M(D^+\to \ov K^0K^{*+}) &=&\lambda_d (A_P+P_P^d)+\lambda_s (1.25\,T_P+P_P^s)-\lambda_b \PE_P, \non \\
\M(D^+\to K^+\ov K^{*0}) &=&\lambda_d (A_V+P_V^d)+\lambda_s (1.22\,T_V+P_V^s)-\lambda_b \PE_V,
\en
with $\M=\tilde \M(\epsilon\cdot p_D)$.

Numerically, we obtain
\be
\begin{split}
\B(D^+\to \ov K^0K^{*+}\to K^+\ov K^0\pi^0) &= 4.92\times 10^{-3}
~, \\
\B(D^+\to K^+\ov K^{*0}\to K^+\ov K^0\pi^0) &= 1.81\times 10^{-3}
~,
\end{split}
\en
and
\be
\begin{split}
a_{C\!P}(D^+\to \ov K^0K^{*+}\to K^+\ov K^0\pi^0) &= 0.06\times 10^{-3}
~, \\
a_{C\!P}(D^+\to K^+\ov K^{*0}\to K^+\ov K^0\pi^0) &= -0.90\times 10^{-3}
~.
\end{split}
\en
The \CP asymmetry distribution of $D^+\to K^+\ov K^0\pi^0$ in phase space is exhibited in Fig.~\ref{fig:DpKKpi_CP}. Regions with positive and negative $a_{C\!P}$ at the percent level
are clearly seen in the plot.

\subsection{$D_s^+\to K^0\pi^+\pi^0$}
Considering the decay amplitude of $D_s^+\to K^0\pi^+\pi^0$ in the overlapped region of the $K^{*+}$ and $K^{*0}$ resonances, we have
\be
A(D_s^+\to K^0(p_1)\pi^+(p_2)\pi^0(p_3))\approx A_{K^{*+}}+A_{K^{*0}},
\en
with
\be
\begin{split}
A_{K^{*+}}
&\equiv A(D_s^+\to \pi^0K^{*+}\to K^0 \pi^+\pi^0)
\\
&= g^{K^{*+}\to K^0\pi^+}F(s_{12},m_{K^*}){  \epsilon^*\cdot(p_2-p_1)\over s_{12}-m_{K^*}^2+im_{K^*}\Gamma_{K^*}(s_{12}) }\M(D_s^+\to \pi^0K^{*+})
~, \\
A_{K^{*0}}
&\equiv A(D_s^+\to \pi^+K^{*0}\to K^0 \pi^+\pi^0)
\\
&= g^{K^{*0}\to K^0\pi^0} F(s_{13},m_{K^*}){  \epsilon^*\cdot(p_1-p_3)\over s_{13}-m_{K^*}^2+im_{K^*}\Gamma_{K^*}(s_{13}) }\M(D_s^+\to \pi^+K^{*0})
~,
\end{split}
\en
where use of the $K^* K\pi$ interaction Lagrangian (\ref{eq:Lag KstKpi})
has been made and
\be
\M(D_s^+\to \pi^0K^{*+}) &=& {1\over\sqrt{2}}\left[ \lambda_d(0.97\,C_V-P_V^d)-\lambda_s(A_V+P_V^s)+\lambda_b\, \PE_V\right], \non \\
\M(D_s^+\to \pi^+K^{*0}) &=&  \lambda_d(0.74\,T_V+P_V^d)+\lambda_s(A_V+P_V^s)-\lambda_b\, \PE_V,
\en
with the SU(3) breaking effects in $T_V$ and $C_V$ included (cf. Table \ref{tab:CSPV}).
Hence,
\be
\begin{split}
A_{K^{*+}}
&= -{1\over 2}\left(s_{23}-s_{13}+{(m_{D_s}^2-m_\pi^2)(m_K^2-m_\pi^2)\over s_{12}}\right) g^{K^{*+}\to K^0\pi^+}
\\
& \qquad\qquad \times F(s_{12},m_{K^*}) T_{K^*}^{\rm BW}(s_{12})\tilde\M(D_s^+\to \pi^0K^{*+})
~, \\
A_{K^{*0}}
&= {1\over 2}\left(s_{23}-s_{12}+{(m_{D_s}^2-m_\pi^2)(m_K^2-m_\pi^2)\over s_{13}}\right) g^{K^{*0}\to K^0\pi^0}
\\
& \qquad\qquad \times F(s_{13},m_{K^*})T_{K^*}^{\rm BW}(s_{13})\tilde\M(D_s^+\to \pi^+K^{*0})
~.
\end{split}
\en

%====================================================================
\begin{figure}[t]
\begin{center}
\centering
 {\includegraphics[scale=0.59]{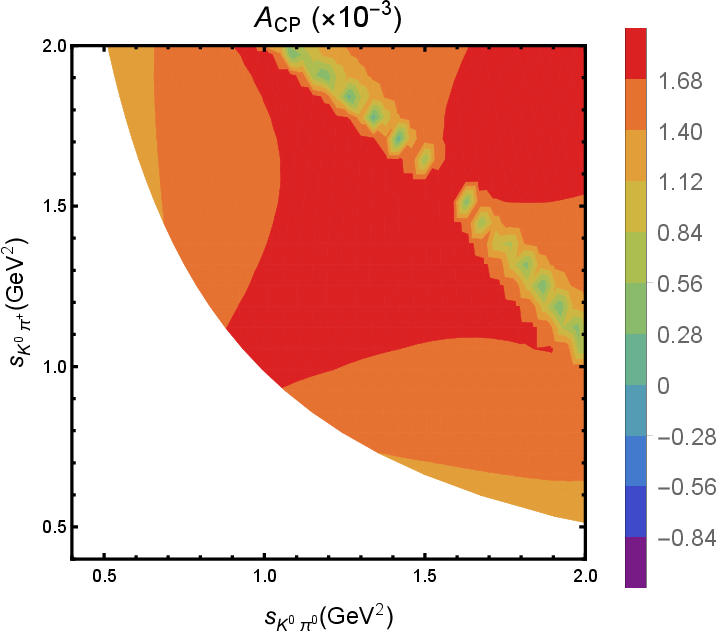}
}
\vspace{0.1cm}
\caption{\CP asymmetry distribution of $D_s^+\to K^0\pi^+\pi^0$ in the overlapped regions of $K^*(892)^+$ and $K^*(892)^0$.}
\label{fig:DsKpipi_CP}
\end{center}
\end{figure}
%=====================================================================

We obtain
\be
\begin{split}
\B(D_s^+\to \pi^0K^{*+}\to K^0\pi^+\pi^0) &= 0.55\times 10^{-3}
~, \\
\B(D_s^+\to \pi^+K^{*0}\to K^0\pi^+\pi^0) &= 0.64\times 10^{-3}
~,
\end{split}
\en
and
\be
\begin{split}
a_{C\!P}(D_s^+\to \pi^0K^{*+}\to K^0\pi^+\pi^0) &= 1.53\times 10^{-3}
~, \\
a_{C\!P}(D_s^+\to \pi^+K^{*0}\to K^0\pi^+\pi^0) &= 1.47\times 10^{-3}
~.
\end{split}
\en
Fig.~\ref{fig:DsKpipi_CP} shows the Dalitz plot of \CP asymmetry distribution of $D_s^+\to K^0\pi^+\pi^0$. Since \CP violation in $D_s^+\to \pi^0K^{*+}$ and $D_s^+\to \pi^+K^{*0}$ is large and positive, \CP asymmetry in most of the phase space is positive, though it can become negative in some regions. For example, it is of order  of order $-0.35\times 10^{-3}$ at $(s_{K^0\pi^0},s_{K^0\pi^+})=(1.97,1.10)\,{\rm GeV}^2$.

\subsection{Finite-width effects \label{sec:Finite-width effects}}
Since all the framework has been set up, we are ready to discuss the finite-width effects. First of all, we need to show that the relation Eq.~(\ref{eq:NWA}) is valid in the narrow width limit. For illustration we take $D^0\to\pi^+\rho^-\to \pi^+(p_1)\pi^-(p_2)\pi^0(p_3)$ as an example.
Since
\be
s_{12}-s_{13}=4\vec{p}_1\cdot\vec{p_2}=4|\vec{p}_1||\vec{p}_2|\cos\theta,
\en
where $|\vec{p}_1|$ and $|\vec{p}_2|$ are the momenta of $\pi^+$ and $\pi^-$, respectively, in the rest frame of $\pi^-$ and $\pi^0$, and
\be
q &=& |\vec{p}_2|=|\vec{p}_3| ={\sqrt{s_{23}-4m_\pi^2}\over 2 }, \non \\
|\vec{p}_1| &=& \left({(m_D^2-m_\pi^2-s_{23})^2\over 4s_{23}} - m_\pi^2\right)^{1/2}={m_D\over \sqrt{s_{23}}}\,\tilde p_c,
\label{eq: q p3}
\en
with $\tilde p_c$ being the c.m.~momentum of $\pi^+(p_1)$ in the $D^0$ rest frame, then the amplitude $A_{\rho^-}$ in Eq.~(\ref{eq:Arho_2})  can be recast to
\be
A_{\rho^-}= -g^{\rho\to \pi\pi}\,F(s_{23},m_{\rho})T_{\rho}^{\rm GS}(s_{23})2q\cos\theta\, \tilde {\cal M}(D^0\to \pi^+\rho^-){m_D\over \sqrt{s_{23}}}\,\tilde p_c.
\en

The decay rate is given by
\be
\Gamma(D^0\to \pi^+\rho^-\to \pi^+\pi^-\pi^0) &=&
{1\over 2}\,{|g^{\rho\to \pi^+\pi^-}|^2\over(2\pi)^3 32 m_D^3}\int ds_{23}\,ds_{12}  { F(s_{23},m_{\rho})^2(1+D \, \Gamma_\rho^0/m_\rho)^2 \over (s_{23}-m^2_{f_2}-f(s_{23}))^2+m_{\rho}^2\Gamma_{\rho}^2(s_{23})} \non \\
&& \times
4q^2\cos^2\theta|\tilde {\cal M}(D^0\to \pi^+\rho^-)|^2 \,{m_D^2\over s_{23}}\,\tilde p_c^2.
\en
The angular distribution part is given by (see Eq. (4.64) of~\cite{Cheng:2020iwk})
\be
\int_{(s_{12})_{\rm min}}^{(s_{12})_{\rm max}}ds_{12}\cos^2\theta={4\over 3}{m_D\over\sqrt{s_{23}}}q\tilde p_c.
\en
In the narrow width limit, we have~\cite{Cheng:2020iwk}
\be
{m_{\rho}\Gamma_{\rho}(s)(1+D \, \Gamma_\rho^0/m_\rho)^2 \over (s-m^2_\rho-f(s))^2+m_{\rho}^2\Gamma_{\rho}^2(s)} \xlongrightarrow[]{\; \Gamma_{\rho}\to 0 \;}\pi\delta(s-m_{\rho}^2-f(s))\to \pi\delta(s-m_{\rho}^2)
~.
\en
As a result of the $\delta$-function,  $s_{23}\to m_\rho^2$.
We then obtain the desired NWA:
\be \label{eq:factorization}
\Gamma(D^0\to \pi^+\rho^-\to \pi^+\pi^-\pi^0) \xlongrightarrow[]{\; \Gamma_{\rho}\to 0 \;}
 \Gamma(D^0\to \pi^+\rho^-) \B(\rho^-\to \pi^-\pi^0)
 ~,
\en
where use of the relations
\be
 \Gamma_{\rho\to \pi^+\pi^-} = {q_0^3\over 6\pi m_{\rho}^2}g_{\rho\to \pi^+\pi^-}^2
 ~, \qquad
 \Gamma_{D^0\to \pi^+\rho^-} = {p_c^3\over 8\pi m_\rho^2}|\tilde\M(D^0\to \pi^+\rho^-)|^2
 ~,
\en
has been made with $q_0$ and $p_c$ being $q$ and $\tilde p_c$, respectively, except for the replacement of $s_{23}$ by $m_\rho^2$. Numerically, we have also verified the NWA (\ref{eq:factorization}).

The finite-width effect is accounted for by the quantity $\eta_R$ defined by \cite{Cheng:2020mna,Cheng:2020iwk}
\be \label{eq:eta}
\eta_{_R}\equiv \frac{\Gamma(D\to RP_3\to P_1P_2P_3)_{\Gamma_R\to 0}}{\Gamma(D\to RP_3\to P_1P_2P_3)}=\frac{\Gamma(D\to RP_3)\B(R\to P_1P_2)}{\Gamma(D\to RP_3\to P_1P_2P_3)}=1+\delta
~,
\en
so that the deviation of $\eta_{_R}$ from unity measures the degree of departure from the NWA when the resonance width is finite.  It is na{\"i}vely expected that the correction $\delta$ will be of order $\Gamma_R/m_R$. The calculated $\eta_V$ parameters for vector resonances $\rho$ and $K^*$ produced in the three-body $D$ decays are summarized in Table~\ref{tab:eta}. Since $\rho$ is three times broader than that of $K^*(892)$, it is na{\"i}vely expected that the deviation of $\eta_\rho$ from unity should be larger than that of $\eta_{K^*}$. However, our calculation shows that they are close to each other.
Our results are to be compared with $\eta_\rho=0.93$ in the GS line shape model and $1.11$ in the BW line shape for $B^+\to \rho^0\pi^+\to \pi^+\pi^-\pi^+$, and $\eta_{K^*}=1.067\pm0.002$ for $B^+\to K^{*0}\pi^+\to K^+\pi^+\pi^-$ \cite{Cheng:2020mna,Cheng:2020iwk}.

%%%%%%%%%%%%%%%%%%%%%%%%%%%%%%%%%%%%%%
\begin{table}[t]
\caption{A summary of the $\eta_R$ parameter for vector resonances produced in the three-body $D$ decays. The parameter $\eta_{\rho^+}$  extracted from $D^0\to \pi^-\rho^+\to \pi^+\pi^-\pi^0$ is the same as $\eta_{\rho^-}$ and likewise for $\eta_{\rho^0}$.
}
\vskip 0.15cm
\label{tab:eta}
\footnotesize{
\begin{ruledtabular}
\begin{tabular}{ l l c c l }
 Resonance~~~ & ~$D\to Rh_3\to h_1h_2h_3$ ~~~ & ~$\Gamma_R$ (MeV)~\cite{PDG}~~ & $\Gamma_R/m_R$ & ~~~$\eta_R$ \\
\hline
$\rho(770)$ & $D^0\to \pi^+\rho^-\to \pi^+\pi^-\pi^0$ & ~$149.1\pm0.8$~~ & 0.192 & ~~0.944 (GS) \\
 &  &  & & ~~1.116 (BW) \\
$K^*(892)$ & $D^0\to K^+ K^{*-}\to K^+K^-\pi^0$ & ~$47.3\pm0.5$~~ & 0.053 & ~~1.112 (BW) \\
$K^*(892)$ & $D^+\to K^+ \ov K^{*0}\to K^+K^-\pi^0$ & ~$47.3\pm0.5$~~ & 0.053 & ~~1.112 (BW) \\
$K^*(892)$ & $D_s^+\to K^{*+}\pi^0\to K^0\pi^+\pi^0$ & ~$47.3\pm0.5$~~ & 0.053 & ~~1.099 (BW) \\
\end{tabular}
\end{ruledtabular} }
\end{table}
%%%%%%%%%%%%%%%%%%%%%%%%%

When the resonance is sufficiently broad, it is necessary to take into account the finite-width effects characterized by the parameter $\eta_R$. Explicitly,
\be
\B(D\to V\!P)=\eta_V\B(D\to V\!P)_{\rm NWA}
~,
\en
where $\B(D\to R P)_{\rm NWA}$ denotes the branching fraction obtained from Eq.~(\ref{eq:NWA}) valid in the NWA. Take $D^0\to \pi\rho$ as an example. Their branching fractions have been extracted from $D^0\to \pi^+\pi^-\pi^0$ by BaBar using the Breit-Wigner line shape~\cite{BaBar:D0pippimpi0}. According to our calculation, $\eta_\rho^{\rm BW}=1.116$. Hence,
the PDG values of $(5.15\pm0.25),(10.1\pm0.4)$ and $(3.86\pm0.23)$ (in units of $10^{-3}$)~\cite{PDG} for the branching fractions of
$D^0\to \pi^+\rho^-,\pi^-\rho^+$ and $\pi^0\rho^0$, respectively, should be corrected to $(5.75\pm0.28),(11.3\pm0.5)$ and $(4.31\pm0.26)$.

\section{Conclusions \label{sec:summary}}

In this work we have re-examined direct \CP violation in the quasi-two-body $D\to V\!P$ decays and studied \CP asymmetries in three-body $D$ decays proceeding through intermediate vector resonances within the framework of topological amplitude approach for tree amplitudes and the QCD factorization approach for penguin amplitudes.
As we have pointed out in 2012, the long-distance penguin-exchange through final-state rescattering gives the major direct \CP violation to both $D^0\to K^+K^-$ and $D^0\to \pi^+\pi^-$. It accounts for nicely the \CP asymmetry difference between the aforementioned two modes observed by the LHCb in 2019. The same mechanism implies that \CP asymmetry can occur at the per mille level in many of the $D\to V\!P$ channels.

 Our main results are:

\begin{itemize}
\item
In light of several new measurements of Cabibbo-favored modes, we have performed a re-fit to the data to extract topological tree amplitudes.  The topological $W$-exchange $E_V$ has a size twice larger than the old value and its phase is significantly different from the old one.

\item
There are six golden modes with sufficiently large branching fractions and \CP asymmetries of order $10^{-3}$: $D^0\to \pi^+\rho^-, K^+K^{*-}$, $D^+\to \eta\rho^+, K^+\overline{K}^{*0}$ and
$D_s^+\to \pi^+ K^{*0}, \pi^0K^{*+}$.  In particular, we predict $a_{CP}(K^+K^{*-})-a_{CP}(\pi^+\rho^-) = (-1.61 \pm 0.33) \times 10^{-3}$, very similar to the observed \CP asymmetry difference between $D^0\to K^+K^-$ and $D^0\to \pi^+\pi^-$ by the LHCb.

\item
In the $U$-spin limit, there are the relations
$a^{\rm dir}_{CP}(D^0\to K^\pm K^{*\mp})= -a^{\rm dir}_{CP}(D^0\to \pi^\pm\rho^\mp)$. We have found that the $U$-spin relation is approximately respected for $D^0\to K^+K^{*-}$ but not for $D^0\to K^-K^{*+}$.

\item
Analogous to the $P\!P$ sector, SU(3) breaking in $E_{V,P}$ is obtained from a fit to eight singly Cabibbo-suppressed $D^0$ decays. There are four different solutions, which can be discriminated by the decay mode $D^0\to \eta\phi$ to favor solution (iv).  An important consequence is that $D^0\to K_S K^{*0}$ is predicted to have a positive \CP asymmetry at the per mille level. It is an example that the asymmetry is induced at the tree level.

\item
The consideration of SU(3)-breaking effects in the $V\!P$ sector is more subtle than the $P\!P$ case. It appears that SU(3) breaking can only occur in one of the topological tree amplitudes in each decay channel. We are able to account for the recent new measurements from BESIII except for $D^+\to K^+\ov K^{*0}$. For this, we need to wait for further experimental justification.

\item
We have included the flavor-singlet QCD-penguin contributions, which were missing in our previous studies, to calculate their effects on modes such as $D\to\pi\phi$ and $D\to \eta \phi$.  As shown in Table~\ref{tab:CPVP}, the inclusion of the singlet penguins in these decay modes results in nonzero $a_{\rm dir}$, albeit not to a level that can be easily measured experimentally.

\item
We compare our approach with the FAT approach in detail. The main difference lies in the treatment of the factorizable part of the penguin-exchange amplitude $\PE$ characterized by $A_3^f$ and the consideration of long-distance contribution  to $\PE$ through final-state rescattering. In QCDF, $A_3^f$ is evaluated in terms of twist-2 and -3 light-cone distribution amplitudes and it does not have near cancellation with $\PE$ as claimed by the FAT analysis.

\item
For three-body $D$ decays, we show the Dalitz plot of \CP asymmetry distribution of $D^0\to K^+K^-\pi^0$ in the overlapped regions of $K^{*+}$ and $K^{*-}$ resonances and likewise for $D^0\to \pi^+\pi^-\pi^0$, $D^+\to K^+K_S\pi^0$ and $D_s^+\to K^0\pi^+\pi^0$. Regional asymmetry varies in magnitude and sign from region to region and can reach the percent level in certain invariant mass regions.
\item
We have considered finite-width corrections to the narrow width approximation for the $\rho(770)$ and $K^*(890)$ resonances. As an example of the implications, the PDG value of $(5.15\pm0.25)\times 10^{-3}$ for the branching fraction of
$D^0\to \pi^+\rho^-$ should be corrected to $(5.75\pm0.28)\times 10^{-3}$.

\end{itemize}

%%%%%%%%%%%%%%%%%%%%%%%%%%%%%%%%%%%%%%%%%%%%%%%%%%
\section*{Acknowledgments}
%%%%%%%%%%%%%%%%%%%%%%%%%%%%%%%%%%%%%%%%%%%%%%%%%%

This research was supported in part by the Ministry of Science and Technology of R.O.C. under Grant Nos.~MOST-107-2119-M-001-034 and MOST-108-2112-M-002-005-MY3.

%%%%%%%%%%%%%%%%%%%%%%%%%%%%%%%%%%%%%%%%%%%%%%%%%%

\end{document}